\documentstyle[psfig,epsf,here]{article}

\def\etal{{\it et~al.}}
\def\GeV{\ifmmode {\mathrm{\ Ge\kern -0.1em V}}\else
                   \textrm{Ge\kern -0.1em V}\fi}%
\def\TeV{\ifmmode {\mathrm{\ Te\kern -0.1em V}}\else
                   \textrm{Te\kern -0.1em V}\fi}%

\newcommand{\Zo}{{\rm Z}}

\newcommand{\qqll}{{\rm q \bar q\ell^+\ell^- }}

\newcommand{\epem}{{\rm e^+e^- }}
\newcommand{\eeZZto}{{\rm e^+e^- \rightarrow ZZ\rightarrow}}

\newcommand{\qqnn}{{\rm q\bar q\nu\bar\nu}}
\newcommand{\qqqq}{{\rm q\bar qq^\prime\bar{q}^\prime}}

\begin{document}

\begin{center}
{\Large \bf Recent Results from the L3 Experiment
\footnote{Invited talk at the XIV International Workshop on Quantum Field Theories 
and High Energy Physics, QFTHEP99, Moscow, Russia, 27 May -- 2 June, 1999, to appear in the proceedings.}} \\

\vspace{4mm}

Salvatore Mele\footnote{
  On leave of absence from INFN--Sezione di Napoli,
  Italy. E--mail: Salvatore.Mele@cern.ch}\\
EP Division, CERN,
CH1211, Gen\`eve 23, Switzerland\\
{\it On behalf of the L3 Collaboration}
\end{center}

%

\begin{abstract}
A data sample corresponding to an integrated luminosity of  232\,pb$^{-1}$ was collected in 1997 and 1998 by the L3 experiment at LEP in $\epem$ collisions at
centre--of--mass energies between $181.7\GeV$ and
$188.7\GeV$. Pair production of fermions and bosons is studied and
compared with the Standard Model expectations. Events with a single
detected photon or W boson are also considered. The
measurement of several Standard Model cross sections is discussed. The
presence and the magnitude of triple couplings of 
charged and neutral electroweak gauge bosons is investigated. These
processes are used to probe New Physics beyond the Standard 
Model, including the existence of extra spatial dimensions.
\end{abstract}

%

%
\section{\bf Introduction}
%

The L3 detector~\cite{l3detector} started in 1989 to collect data at
LEP in $\epem$ collisions at a centre--of--mass energy
$\sqrt{s}\simeq m_\Zo$. It has contributed to the success of the
LEP\,I programme, when more than twenty millions of Z bosons were
produced   testing the Standard Model of
electroweak interactions~\cite{sm} (SM) with an
impressive precision~\cite{mnich}. In 1995 the campaign for the gradual increase
of the LEP beam energy started, bringing the experiments in the so called LEP\,II era. In
the present paper I will give a snapshot of several L3 results from the
high integrated luminosity runs of 
 1997 and 1998 at an average $\sqrt{s}$ of $182.7\GeV$ and 
$188.7\GeV$, respectively. These two
energies will be  indicated as $183\GeV$ and $189\GeV$
hereafter and correspond to 55\,pb$^{-1}$ 
 and 176\,pb$^{-1}$ of integrated luminosity, respectively. Some of the results
described here are published, others are preliminary. Results from the
lower energy runs at $133-140\GeV$ with 
10\,pb$^{-1}$ of integrated luminosity and $161\GeV - 172\GeV$ with
21\,pb$^{-1}$, are sometimes 
included in the discussed analyses.

These results, which are only a subsample of the LEP\,II physics
program carried out by the L3 experiment, are classified in the following
according to the type and number of particles primarily produced in the $\rm
e^+e^-$ interactions: two fermions, two W, just one W, two or more
photons, one Z and a photon and two Z. Tests of theories of gravity
with extra spatial dimensions are finally presented.

%
\section{\bf Two fermions}
%

Fermion pair production is a fundamental process to be studied at
LEP\,II both as a verification of
SM predictions and as a necessary check of  the understanding of 
the detector performance. A
kinematically favoured configuration in fermion pair production at
energies above the Z pole is the emission of an hard initial state photon
which lowers the effective centre--of--mass energy, $\sqrt{s'}$, to the Z
resonance. This process is known as ``radiative return to the Z'' and
yields high energy photons either in the detector or  almost collinear
with the beams and hence undetected.

\begin{figure}[H]
\centerline{\mbox{
\begin{tabular}{cc}
\mbox{\psfig{figure=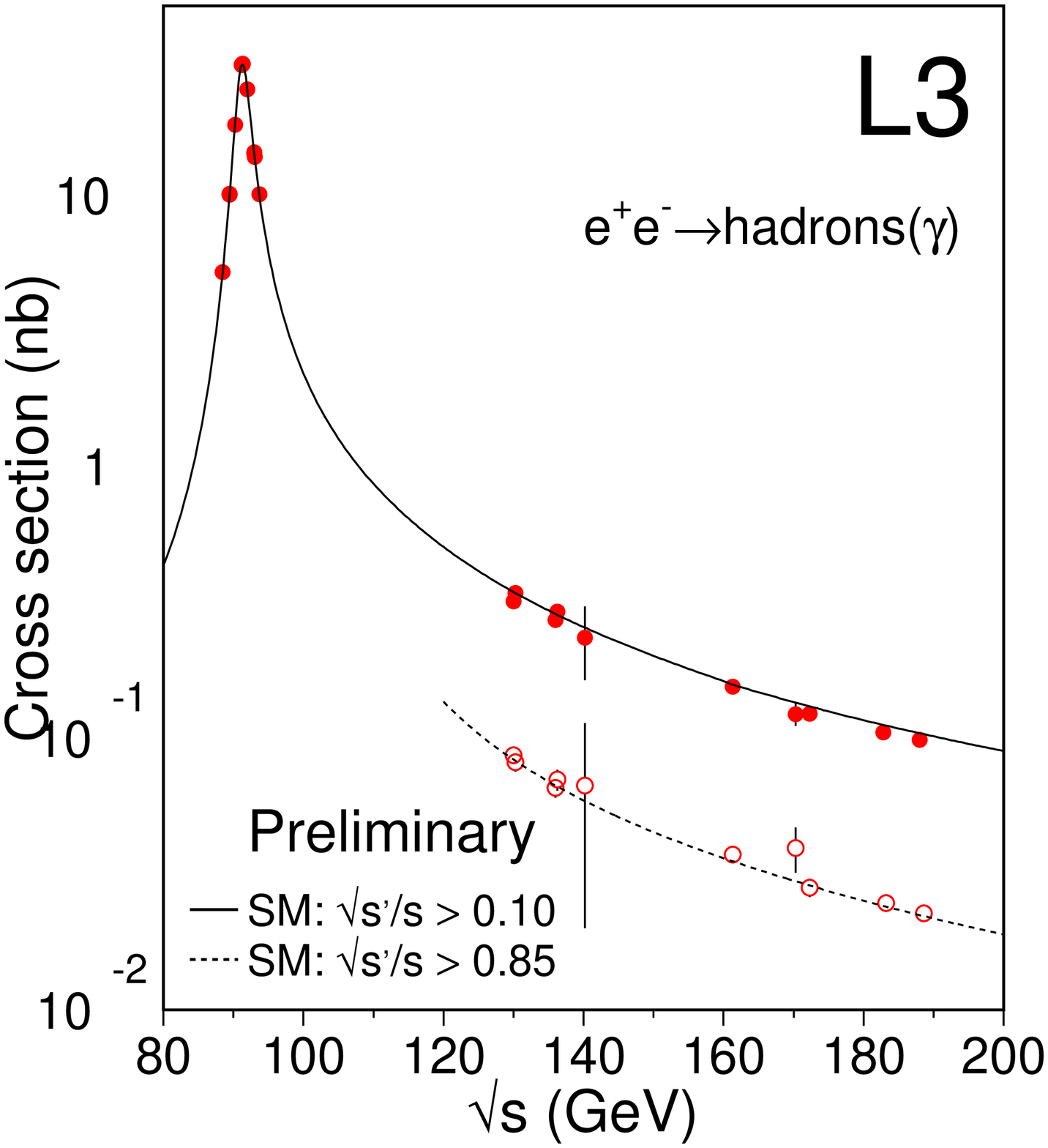,width=0.45\textwidth}} &
\mbox{\psfig{figure=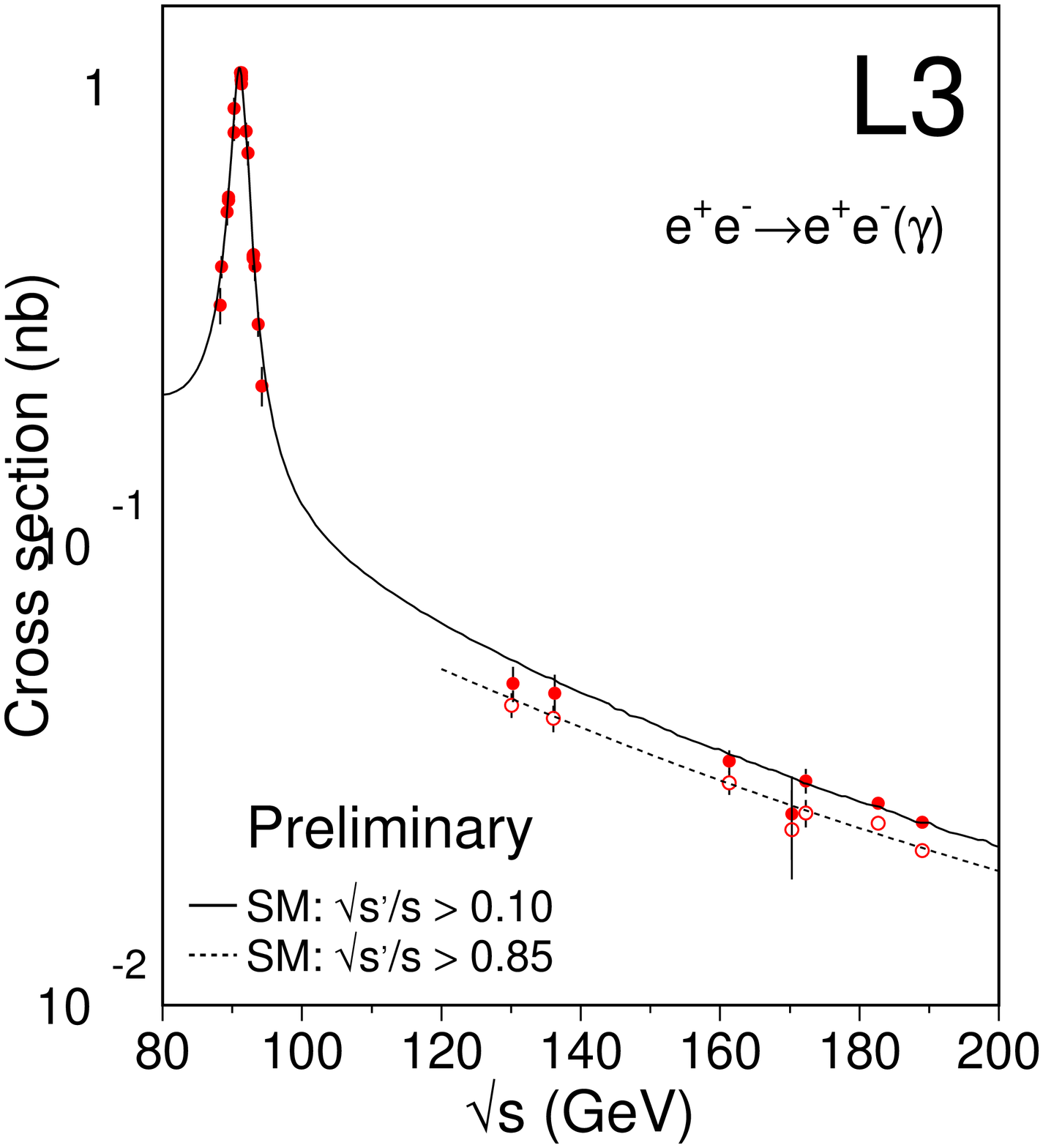,width=0.45\textwidth}} \\
\mbox{\psfig{figure=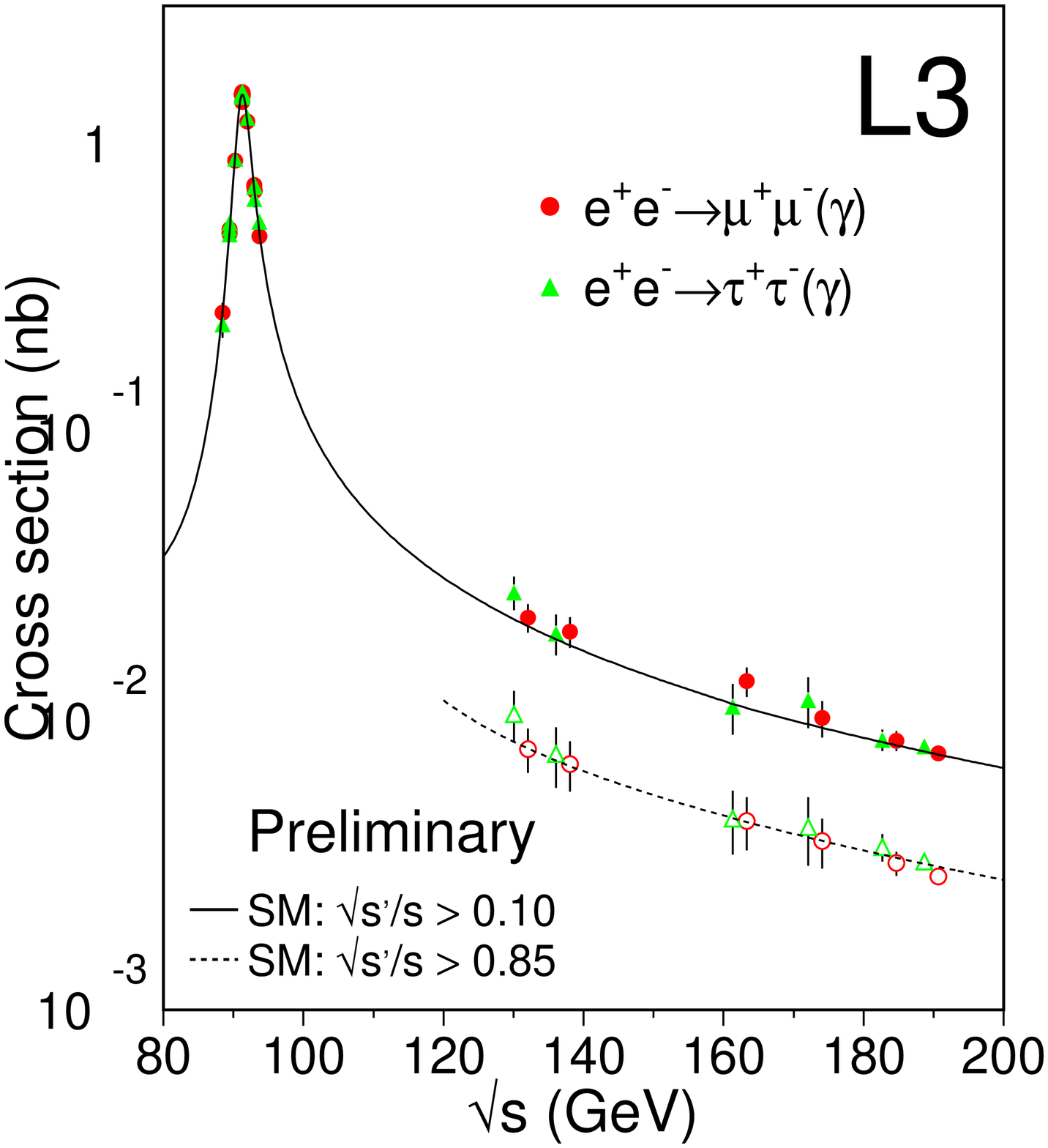,width=0.45\textwidth}} &
\mbox{\psfig{figure=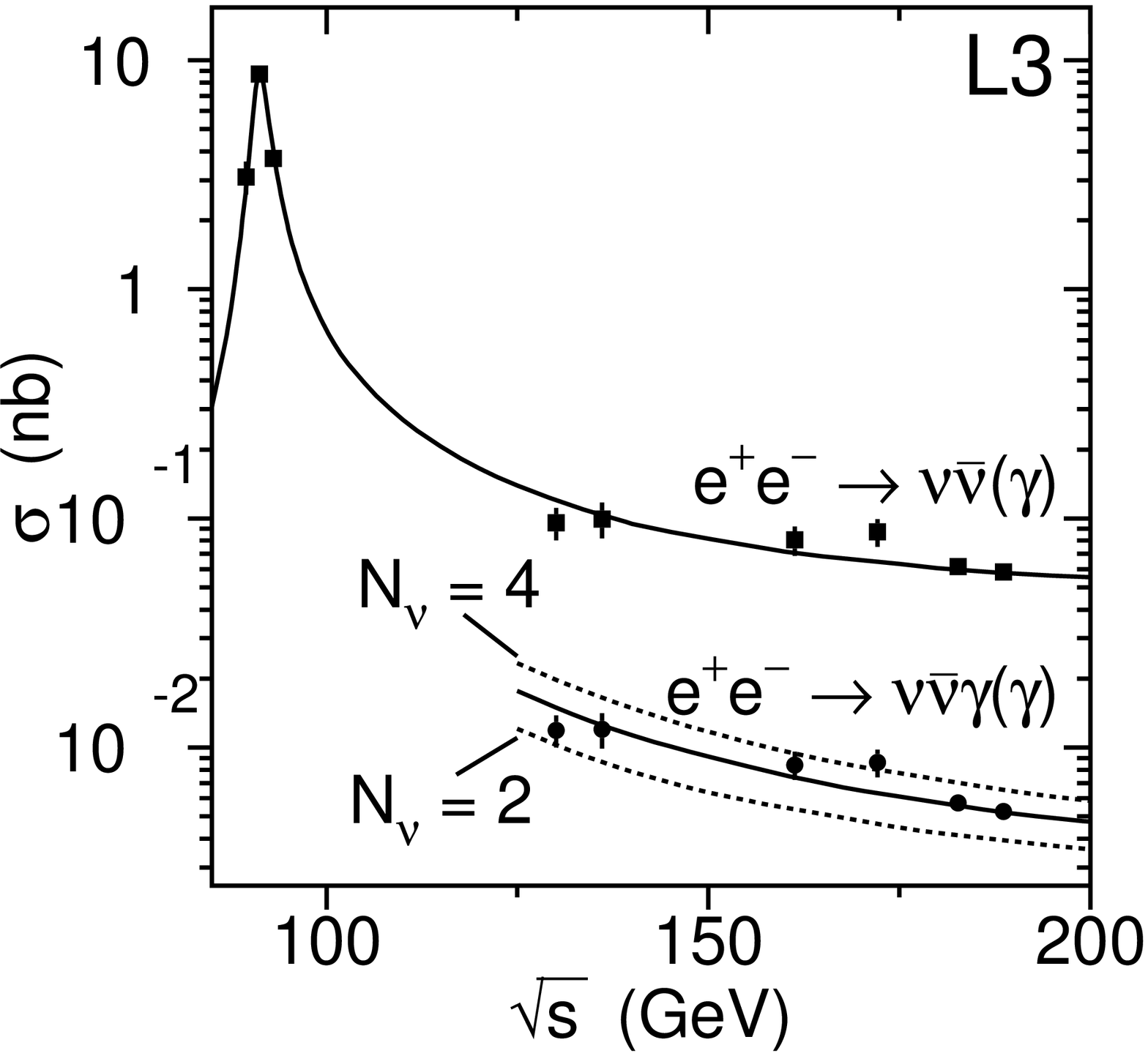,width=0.45\textwidth}} \\
\end{tabular}
}}
\caption{\label{fig:1} Fermion pair cross section evolution as a
  function of the
  centre--of--mass energy.}
\end{figure}

 It is customary to express the
results of the cross sections and forward--backward asymmetries of the
fermions by separating the full data sample ($\sqrt{s'/s}>0.1$) and the
purely high energy one, from which the radiative return to the Z events
are rejected ($\sqrt{s'/s}>0.85$). Table~1 reports the L3  results for
the two considered energies~\cite{fftampere} for quark, muon, tau and
electron pairs. A good agreement with the SM predictions is observed.
Figure~\ref{fig:1} presents the evolution with $\sqrt{s}$ of the fermion pair
cross section as predicted by the SM and measured by the L3 experiment.

From the analysis of events with a single photon visible in the
detector~\cite{l3singlephoton} it is possible to derive the radiative
neutrino pair production cross section. At $189\GeV$ the measured cross section 
is:
\begin{displaymath}
\rm \sigma_{e^+e^-\rightarrow \nu\bar{\nu}\gamma(\gamma)}(189\GeV) = 5.25 \pm 0.22 \pm 0.07\,pb,
\end{displaymath}
where the first error is statistical and the second systematic. The
extrapolation of this value to the total cross section reads:
\begin{displaymath}
\rm \sigma_{e^+e^-\rightarrow \nu\bar{\nu}(\gamma)}(189\GeV) = 58.3\pm2.5\,pb.
\end{displaymath}

\begin{table}[htb]
  \begin{center}
    \begin{tabular}{|l|c|c|c|c|}
      \hline
      & \multicolumn{2}{c|}{Cross sections (pb)} & 
      \multicolumn{2}{c|}{Asymmetries}\\
      \hline
       \multicolumn{5}{|c|}{$\sqrt{s} = 183\GeV$}\\
      \hline
      Channel & $\sqrt{s'/s}>0.1$ & $\sqrt{s'/s}>0.85$ & 
      $\sqrt{s'/s}>0.1$ & $\sqrt{s'/s}>0.85$ \\
      \hline
      $\rm e^+e^-\rightarrow q\bar{q}    (\gamma)$ &$105.9\pm1.5$&$24.2 \pm 0.8$ & -- & -- \\ 
      $\rm e^+e^-\rightarrow\mu^+\mu^-   (\gamma)$ &$8.6\pm0.7  $&$3.2 \pm 0.3$  &$0.38\pm0.07$&$0.56\pm0.07$\\
      $\rm e^+e^-\rightarrow\tau^+\tau^- (\gamma)$ &$8.6\pm0.8  $&$3.7\pm0.4$    &$0.28\pm0.09$&$0.59\pm0.09$\\
      $\rm e^+e^-\rightarrow e^+e^-      (\gamma)$ &$27.5\pm0.7 $&$24.9\pm0.7$   &$0.80\pm0.02$&$0.83\pm0.02$\\
      \hline
       \multicolumn{5}{|c|}{$\sqrt{s} = 189\GeV$}\\
      \hline
      $\rm e^+e^-\rightarrow q\bar{q}    (\gamma)$ &$99.4\pm0.8$&$22.8 \pm 0.4$ & -- & -- \\ 
      $\rm e^+e^-\rightarrow\mu^+\mu^-   (\gamma)$ &$7.7\pm0.3  $&$3.4 \pm 0.2$  &$0.26\pm0.03$&$0.57\pm0.04$\\
      $\rm e^+e^-\rightarrow\tau^+\tau^- (\gamma)$ &$8.0\pm0.4  $&$3.1\pm0.2$    &$0.29\pm0.03$&$0.53\pm0.05$\\
      $\rm e^+e^-\rightarrow e^+e^-      (\gamma)$ &$25.1\pm0.4 $&$21.7\pm0.4$   &$0.82\pm0.01$&$0.85\pm0.01$\\
      \hline
    \end{tabular}
    \caption{Experimental cross sections and asymmetries of  fermion pair production.}
  \end{center}
\end{table}

Figure~\ref{fig:1} shows the evolution of these two  cross sections as
a function of  $\sqrt{s}$  as well as a comparison of $\sigma_{e^+e^-\rightarrow
  \nu\bar{\nu}\gamma(\gamma)}$ with the SM expectations in presence of
2, 3 or 4 neutrino species. From the  photon
energy spectrum  it is possible
to determine the number $N_\nu$ of light neutrino species 
as $N_\nu=3.05\pm 0.11 \pm 0.04$, with respectively statistic and systematic
errors. The average of this result with  L3 results at the Z
pole~\cite{l3nnu} yields:
\begin{displaymath}
N_\nu = 3.011 \pm 0.077.
\end{displaymath}

The study of fermion pairs constitutes also an interesting probe of
possible New Physics beyond the SM. The examples of Supersymmetry and contact
interactions will be highlighted in the following.

\begin{figure}[H]
\centerline{\mbox{
\begin{tabular}{cc}
\mbox{\psfig{figure=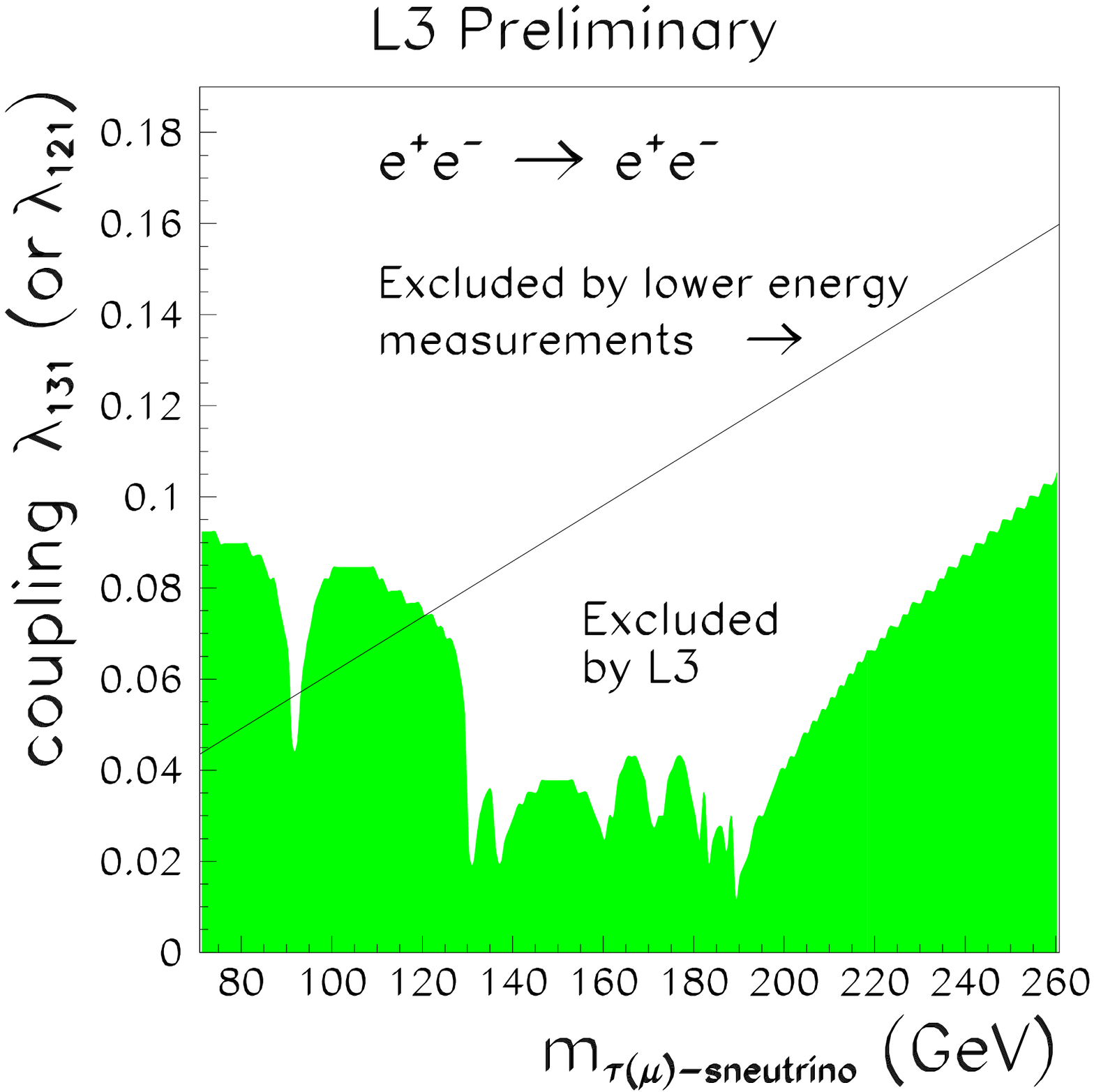,width=0.5\textwidth}} &
\mbox{\psfig{figure=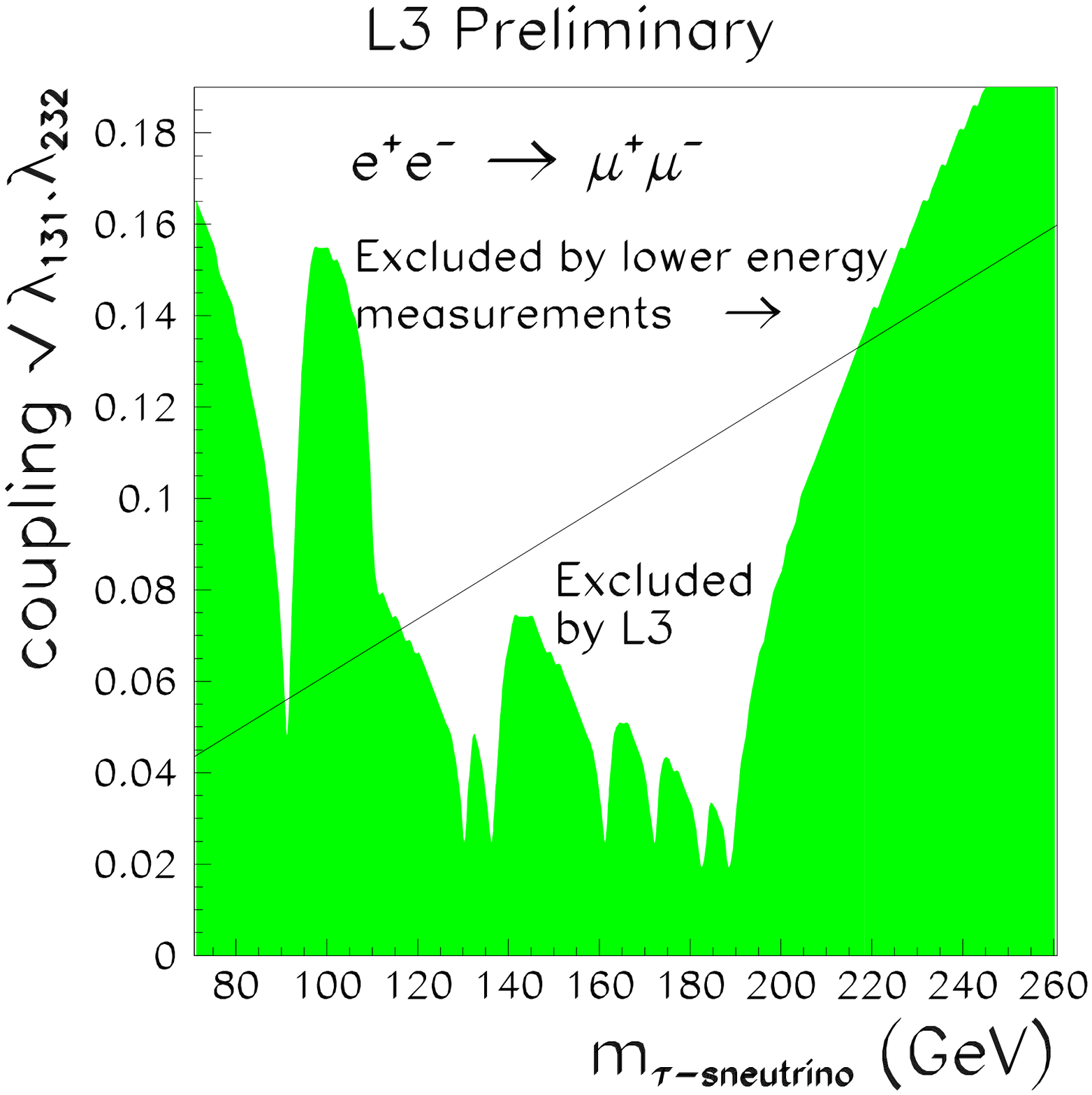,width=0.5\textwidth}} \\
\end{tabular}
}}
\caption{\label{fig:snu} Limits at 95\% CL on the R parity violating
  coupling as a function of the mass of the exchanged sneutrino.}
\end{figure}

The  Minimal Supersymmetric Standard
Model~\cite{mssm} (MSSM) introduces a new quantum number,
the R--parity~\cite{rparity}. This quantity  distinguishes ordinary particles and
supersymmetric ones, requiring an even number of the latter in each
interaction vertex and hence constraining the lightest supersymmetric
particle to be stable. As a consequence of the negative results of the
search for supersymmetric particles at the present colliders, it is
interesting to investigate possible signatures of supersymmetric
models with broken R--parity~\cite{rparitybroken}, where triple
vertices with a supersymmetric particle and two SM ones are allowed. Some of these
signatures are the production of a pair of electrons in $\epem$
collisions mediated by the $s-$ or $t-$channel muon or tau sneutrino
exchange, the production of a muon pair via the $s-$channel
exchange of a tau sneutrino or that of a tau pair via a muon
sneutrino. These sneutrinos are the scalar partner of the SM neutrinos.
The presence of these processes would lead to a resonant structure
in the lepton pair production cross section around
the mass of the exchanged sneutrino. From the investigation~\cite{ffsnu} of the cross
sections and asymmetries of final states electrons, muons and tau, no
evidence for such signatures is found and limits at 95\% confidence
level (CL) are derived on the coupling constant of the sneutrino as a
function of its mass, as reported in Figure~\ref{fig:snu}, for the electron and muon signatures.

\begin{figure}[H]
\centerline{\mbox{
\begin{tabular}{cc}
\mbox{\psfig{figure=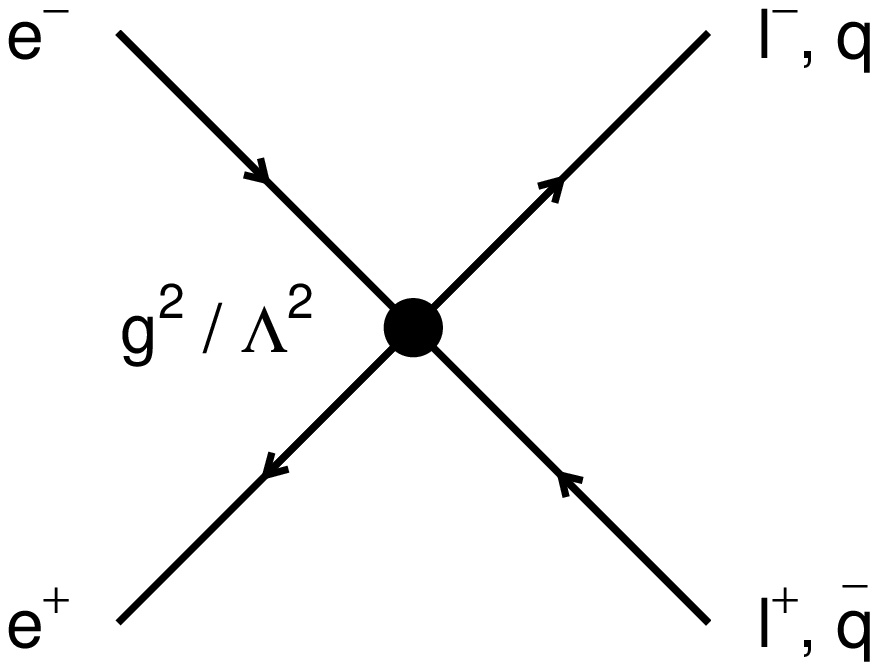,width=0.4\textwidth}} &
\mbox{\psfig{figure=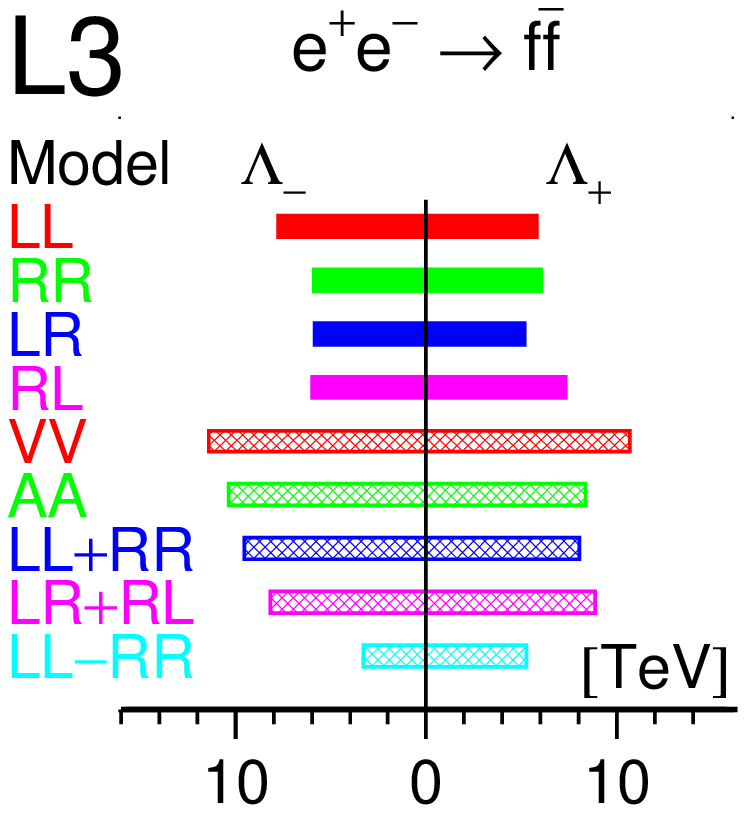,width=0.4\textwidth}} \\
\end{tabular}
}}
\caption{\label{fig:ci} Feynman diagram for the contact interactions
  and 95\% CL limits on their scale in different models.}
\end{figure}

Contact interactions can be thought of as a general formalism to describe
New Physics from a scale much higher than the energy of an investigated
process. An example is the contact interaction structure
 used by Fermi to describe the beta decay~\cite{fermi} fifty years
 before  colliders reached 
the necessary energy to produce the W boson, whose mass is now known
to be the
scale of the process. Analogously a new interaction of coupling
constant $g$ and scale $\Lambda$ yet far above direct experimental reach can be probed in fermion pair production in $\epem$ interactions.
It is sketched in Figure~\ref{fig:ci} and is parametrised via an effective 
Lagrangian~\cite{lagrangian}:

\begin{displaymath}
{\cal{L}} = {1 \over 1+\delta_{\mathrm{e}f}}
\sum_{i,j = {\mathrm{L,R}}}
\eta_{ij} {g^2 \over \Lambda_{ij}^2} 
(\overline{\mathrm{e}_i} \gamma^{\mu} \mathrm{e}_i)
(\overline{f_j} \gamma_\mu f_j),
\end{displaymath}
where $\mathrm{e}_i$ and $f_j$ are the left-- and right--handed
initial state electron and final state fermion fields and the coefficients
$\eta_{ij} = 0, \pm 1$ allow to choose which helicities contribute to the
fermion pair production within the different models, as listed
in Table~2.

    The contact interaction  will manifest itself in the 
differential cross sections of a given phase space parameter
as a
function of  $1/\Lambda^2$ in the interference terms of SM and New
Physics and as a function of  $1/\Lambda^4$ for the pure New Physics part.
The search for contact interactions~\cite{ffnewphys} proceeds by
performing a $\chi^2$ fit to the charged fermion pair cross sections and
asymmetries measurements presented above, with $\Lambda$ as a free parameter
with the convention $g^2/4 \pi =1$. The results of those fits are
compatible with the SM for all the possible choices of the helicities
and 95\% CL limits as high as $12\TeV$ are  set on the scale $\Lambda$
of some models. All the limits  are presented in Figure~\ref{fig:ci},
where  $\Lambda_+$ and $\Lambda_-$ denote respectively the limits in the case of the
upper and lower signs of the $\eta_{ij}$ parameters of Table~2.
Lower energy data~\cite{cilow} are also included.

\begin{table}[H]
  \begin{center}
    \begin{tabular}{|c|ccccccccc|}
      \hline
      Model & LL     & RR   & LR   & RL   & VV   & AA   & LL+RR & LR+RL 
      & LL-RR \\
      \hline
      $\eta_{\mathrm{LL}}$ & $\pm1$ & 0    & 0    & 0    &$\pm1$&$\pm1$& $\pm1$& 0    
      & $\pm1$\\
      $\eta_{\mathrm{RR}}$ & 0      &$\pm1$& 0    & 0    &$\pm1$&$\pm1$& $\pm1$& 0    
      & $\pm1$\\
      $\eta_{\mathrm{LR}}$ & 0      & 0    &$\pm1$& 0    &$\pm1$&$\pm1$& 0     & $\pm1
      $& 0 \\
      $\eta_{\mathrm{LR}}$ & 0      & 0    & 0    &$\pm1$&$\pm1$&$\pm1$& 0     & $\pm1
      $& 0 \\
      \hline
    \end{tabular}
    \caption{Helicities contributions in different models of contact interactions.}
  \end{center}
\end{table}

%
\section{\bf Two W}
%

The study of W pair production constitutes the core of the LEP\,II
physics program and started in 1996 when the centre--of--mass energy of
the LEP machine reached the threshold of $161\GeV$. The SM describes this process at the lowest
order with three diagrams: the $t-$channel neutrino exchange and the
$s-$channel exchange of a Z or a photon. These last two diagrams are
of crucial importance in the SM as they are a manifestation of its non Abelian
structure that allows the triple vertices, ZWW and $\gamma$WW, 
of electroweak gauge bosons.

\begin{figure}[H]
\centerline{\mbox{
\begin{tabular}{cc}
\mbox{\psfig{figure=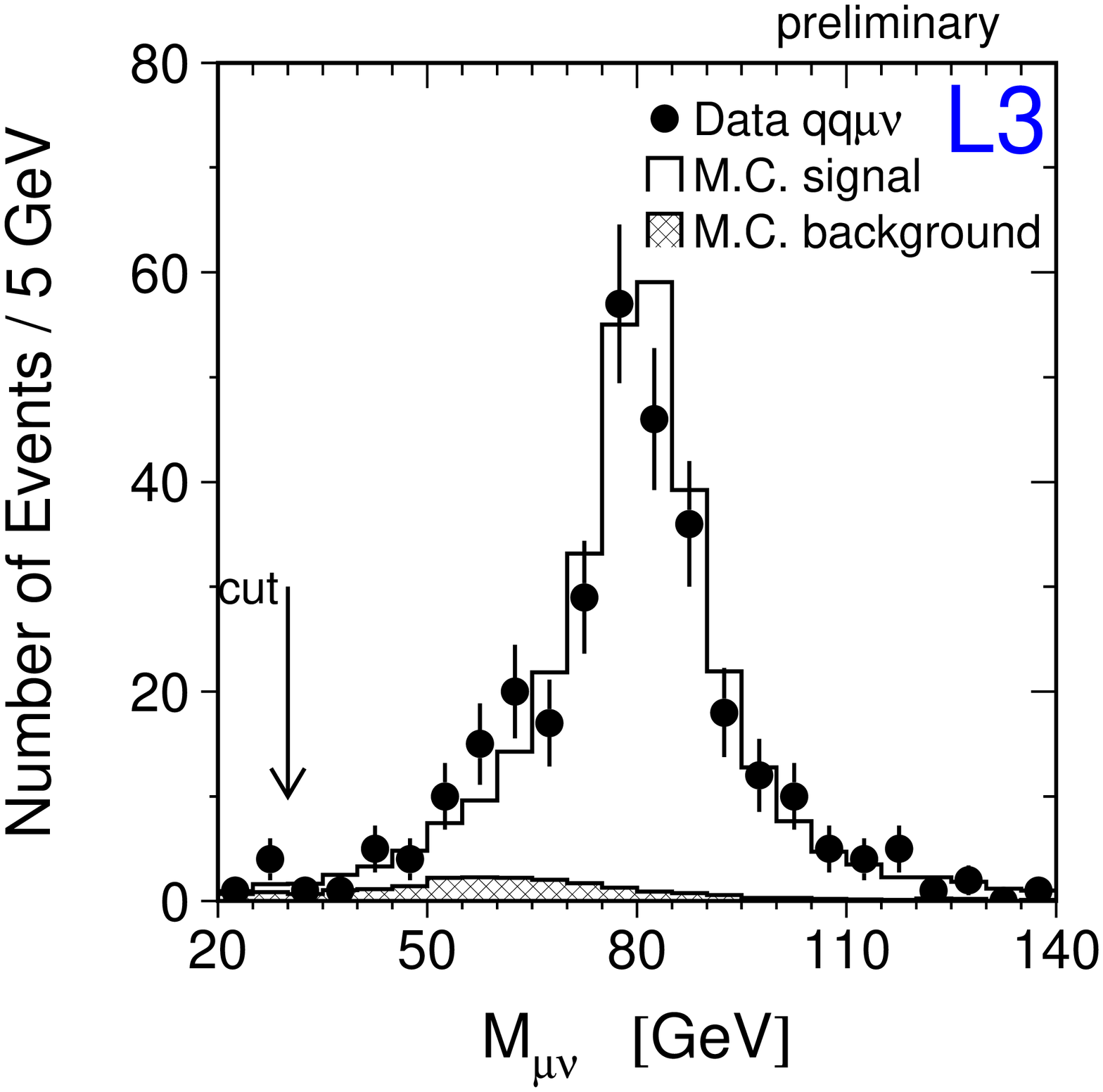,width=0.5\textwidth}} &
\mbox{\psfig{figure=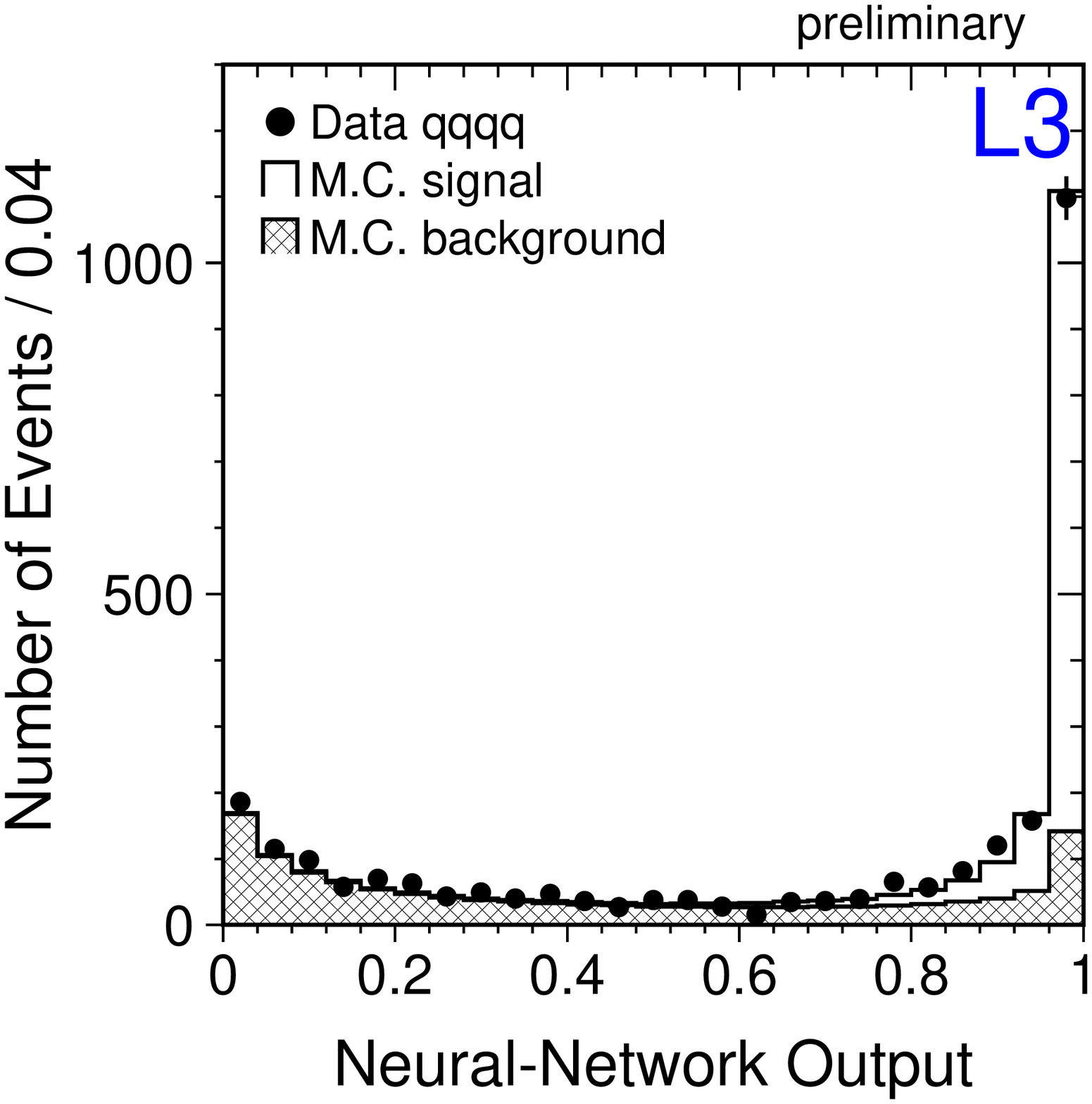,width=0.5\textwidth}} \\
\end{tabular}
}}
\caption{\label{fig:wwsel} Selection variables for W pair events.}
\end{figure}

Detailed descriptions of the combined results of the
four LEP experiments are presented elsewhere~\cite{theseproc}. In
the following, emphasis will be put on results achieved by the L3
experiment. All these results start from a selection of W pair events
in the collected data sample. Five selections are devised to cope with the
fully leptonic decay of the W pair, its semileptonic decay in a quark
pair and  an electron, a muon or a tau together with their
associated neutrinos and finally its fully hadronic decay.
 Examples of selection variables are
reported in Figure~\ref{fig:wwsel} for  semileptonic decays of the W
pair into a muon and its neutrino and fully  
hadronic decays. The latter is selected by means of a neutral
network. A high purity is achieved and combining
all the decay modes, correcting for the contributions to the same
final states from diagrams other than the W pair production,
the cross section of W pair production is measured as~\cite{wwsel}:

\begin{displaymath}
\sigma_{\epem\rightarrow\rm WW}(189\GeV) = 16.25 \pm 0.38 \pm 0.27\rm\,pb,
\end{displaymath}
where the first error is statistical and the second systematic.

\begin{figure}[thb]
\centerline{\mbox{
\begin{tabular}{cc}
\mbox{\psfig{figure=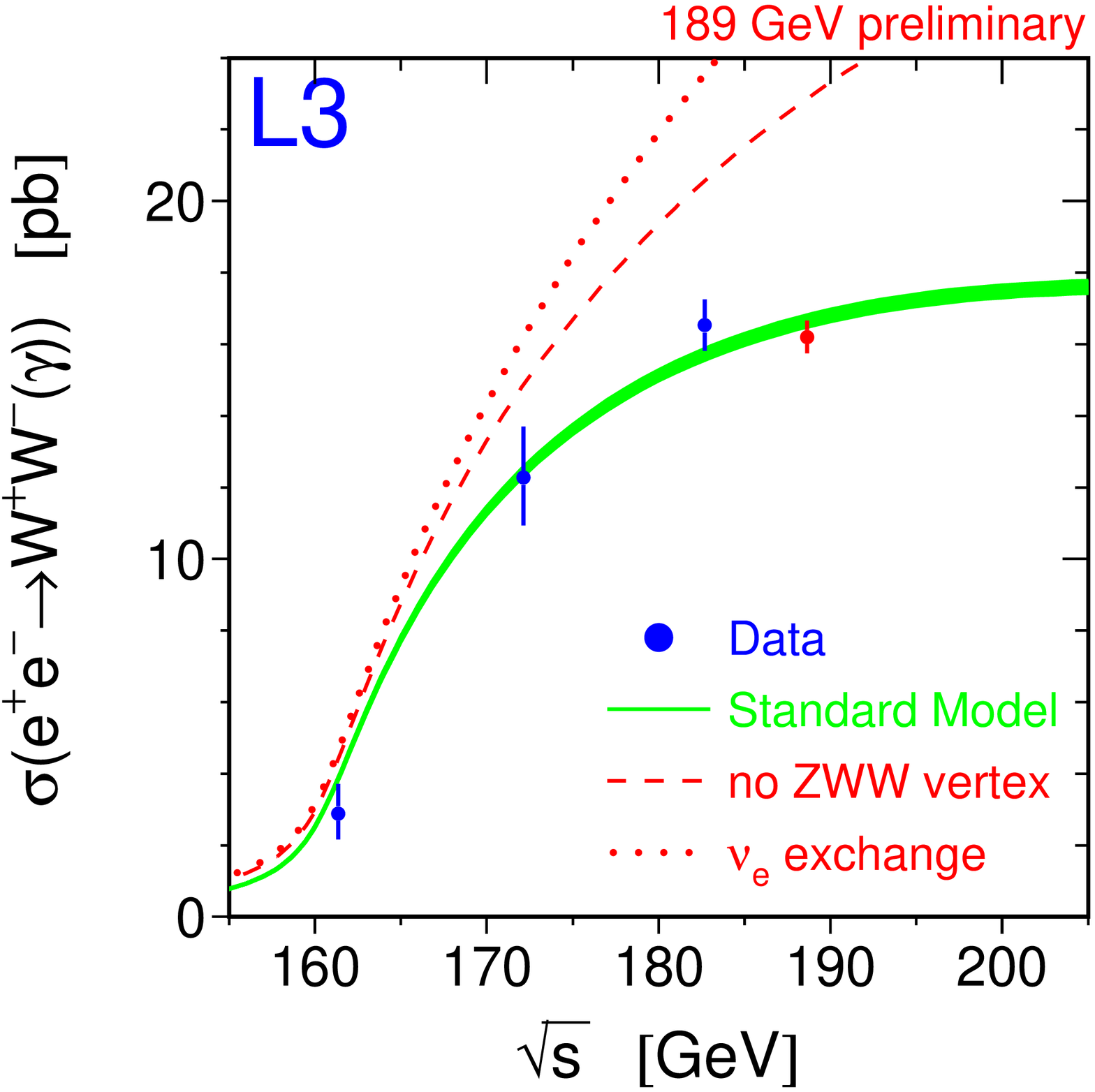,width=0.5\textwidth}} &
\mbox{\psfig{figure=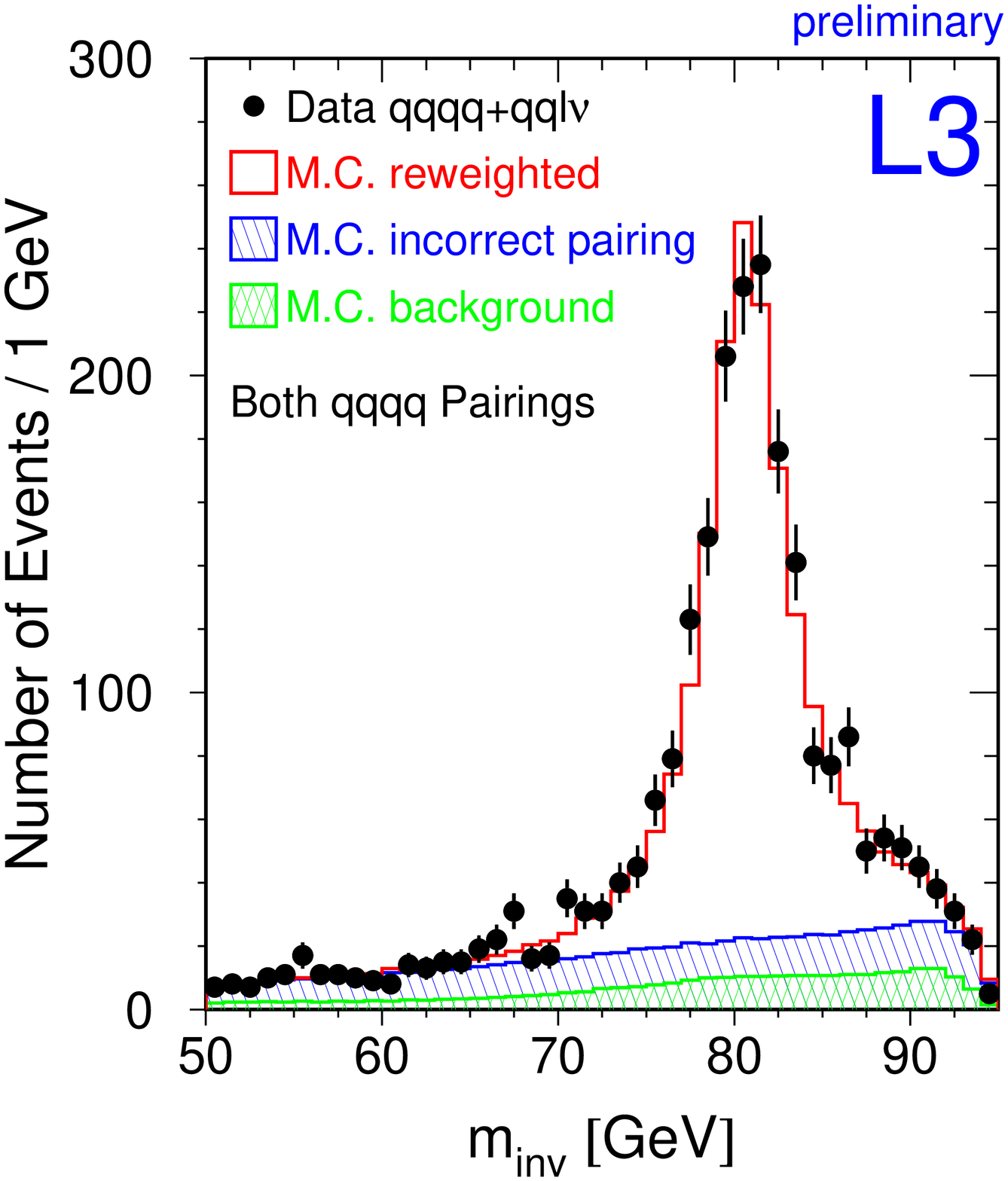,width=0.5\textwidth}} \\
\end{tabular}
}}
\caption{\label{fig:2} Evolution with the centre--of--mass energy of
  the W pair production cross section and distribution of the
  reconstructed W mass  used in the fit.}
\end{figure}

Figure~\ref{fig:2} presents the evolution of the WW cross section with
$\sqrt{s}$ as measured by L3, compared with the theory
predictions. Predictions for W 
pair production via only the electron neutrino $t-$channel exchange
are also reported together with those in absence of the triple gauge
boson vertex ZWW. The evidence of the presence of the $\gamma$WW and
ZWW constitutes an impressive  proof of the non Abelian structure of the SM.
 The measured cross sections are 
 in good agreement with the SM prediction even though this comparison is limited
by the theory uncertainty, as large as 2\%, due to be reduced to about
0.5\% in the near future~\cite{MCworkshop}.

The two vertices  $\gamma$WW and
ZWW are described by means of seven complex coupling
each~\cite{hagiwara}. At energies below the scale of possible New
Physics, only the real part of the couplings is of interest. The simultaneous determination of these
parameters is impossible with the available statistics and their
number is then restricted to five  by first discarding the C--, P-- and
CP--violating ones and assuming electromagnetic gauge invariance.
The extra requirement of custodial SU(2) symmetry allows to further reduce the
number of independent couplings to just three: $g_1^{\rm Z}$, $\kappa_\gamma$ and
$\lambda_\gamma$. It is interesting to relate these three couplings to 
basic physical quantities. $g_1^{\rm Z}$ is the weak coupling strength
of the produced W pair to the exchanged Z boson, while $\kappa_\gamma$ and
$\lambda_\gamma$ enter in the definition of the static magnetic dipole
$\mu_{\rm W}$
and electric quadrupole $Q_{\rm W}$ momenta of the W boson:
\begin{displaymath}
\mu_{\rm W} = {e \over 2m_{\rm W}}(1 + \kappa_\gamma +
\lambda_\gamma); \,\,\,\,
Q_{\rm W} = {e \over 2m_{\rm W}}(1 + \kappa_\gamma - \lambda_\gamma).
\end{displaymath}

Apart from the total cross section measurement, 
more information on these couplings is provided by the distribution of
the polar angle of the produced W bosons and their fermion decay
angles. From an analysis of hadronic and semileptonic decays of 
the W pair 
the  values of these couplings are measured as~\cite{tgc}:
\begin{displaymath}
g_1^{\rm Z} = 0.98 \pm 0.07 \pm 0.03;\,\,\,\,
\kappa_\gamma = 0.88 ^{+0.14} _{-0.12} \pm 0.08;\,\,\,\,
\lambda_\gamma = 0.00  \pm 0.07 \pm 0.03,
\end{displaymath}
where the first error is statistical and the second systematic.
These values are in perfect agreement with the expected SM values of
1, 1 and 0, respectively. This agreement is also found if
two-- or three--parameter fits are performed.

The selected events in each of the semileptonic and hadronic decay
modes of the W pairs are used separately to determine the mass and the width of the W
boson. Kinematic constraints are applied on the events and the
weighted average mass $M_{inv}$ is constructed for each of them. The value of
this mass in data is compared to reweighted Monte Carlo (MC) events with a complex
fit algorithm~\cite{wmastampere} to measure the W
mass. Figure~\ref{fig:2} presents the distribution of $M_{inv}$ for
data collected at $189\GeV$ and the fit MC. The average of the result
of the fit procedure to all the W pair decay channels at $172\GeV$~\cite{wmas172}, $183\GeV$~\cite{wmas183}
and  $189\GeV$, including the mass measurement from the threshold
production cross section at $161\GeV$~\cite{w161} and $172\GeV$~\cite{w172} is:
\begin{displaymath}
M_{\rm W} = 80.43 \pm 0.11 \GeV,
\end{displaymath}
in agreement with the SM expectation observed from a combined fit to the
available data~\cite{mnich}. If the W width is left free in these fits, its value  is
measured in the previous data sample as:
\begin{displaymath}
\Gamma_{\rm W} = 2.12 \pm 0.25 \GeV.
\end{displaymath}

%
\section{\bf Just one W}
%

In the fall of 1996 LEP was running at $\sqrt{s}=172\GeV$ and an
unexpected number of acoplanar hadronic  event with large missing
energy was observed by the L3 analyses surveying the data 
for New Physics signatures. Contrarily
to our hopes these events had a poor content of b
quarks, and hence were incompatible with the hypothesis of an Higgs
boson produced in association to a Z decaying into neutrinos. The first single
W events were observed, as expected~\cite{yb}. This was confirmed by counting  events with a
nothing but a single lepton in the detector to be in the relative
amount with respect to the number of hadronic events as expected from the
W branching ratios~\cite{singleW1}. 

Even though a discovery was missed, sound physics information is
 extracted from this kind of events. The so called single W signal is a subset of the 20
diagrams that lead to the production of four fermions in the final
state, two of which come from the decay of a W, the other two being an
electron and its neutrino. The collective
name of CC20 indicates this $\epem\rightarrow \rm e^+\nu_{\rm e} f
\bar{f}'$ process and its charge conjugate. In the case of interest, the process occurs
via the $t-$channel exchange of a virtual W from the incoming electron
or positron and a virtual photon from the other incoming particle,
giving  a real W via a WW$\gamma$ vertex. The cross section
for single W production is strongly peaked for almost unscattered
electrons or positrons and presents several challenges in its
calculation~\cite{MCworkshop}. This behaviour requires the introduction
of some phase space cuts for the definition of the signal:
\begin{displaymath}
|\cos{\theta_{\rm e^+}}| > 0.997;\,\,\,
\rm min(E_f,E_{f'}) > 15\GeV;\,\,\,
|\cos{\theta_{\rm e^-}}| > 0.75\,\,\,\, (\rm for \,\,\,\rm e^+\nu_{\rm e}e^-\bar{\nu_{\rm e}}).
\end{displaymath}
The angles refer to the polar angles with respect to the beam line and
charge conjugate particles are also included. $E_f$ and $E_{f'}$ are the energies
of the fermions. The application of this signal definition to the CC20
processes yields a single W sample with a purity of 90\%.

The presence of only the WW$\gamma$
vertex makes this process suitable for an accurate study of the
electromagnetic couplings of the W boson, namely its static magnetic dipole
$\mu_{\rm W}$ and electric quadrupole $Q_{\rm W}$, introduced above.

\begin{figure}[thb]
\centerline{\mbox{
\begin{tabular}{cc}
\mbox{\psfig{figure=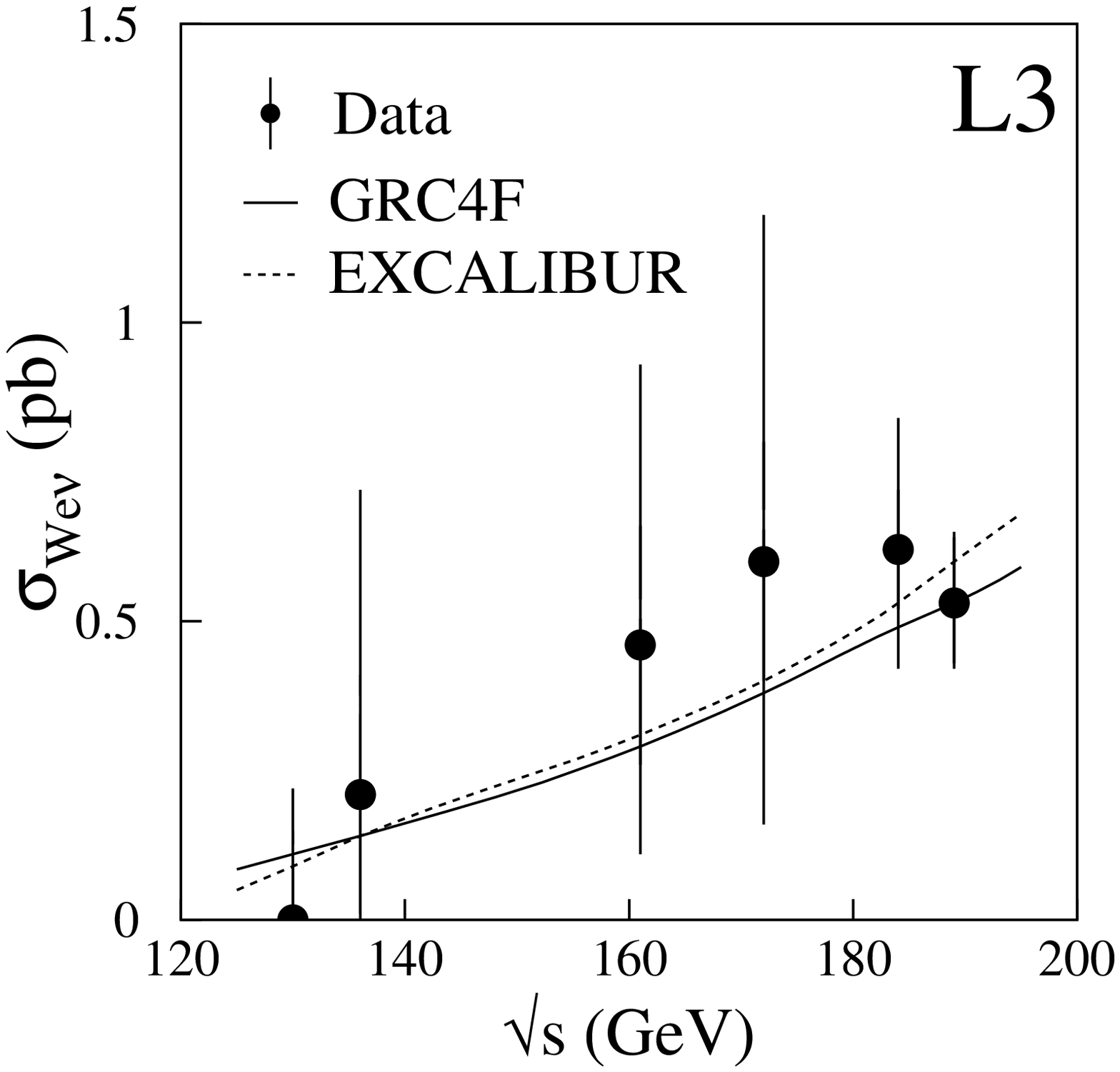,width=0.5\textwidth}} &
\mbox{\psfig{figure=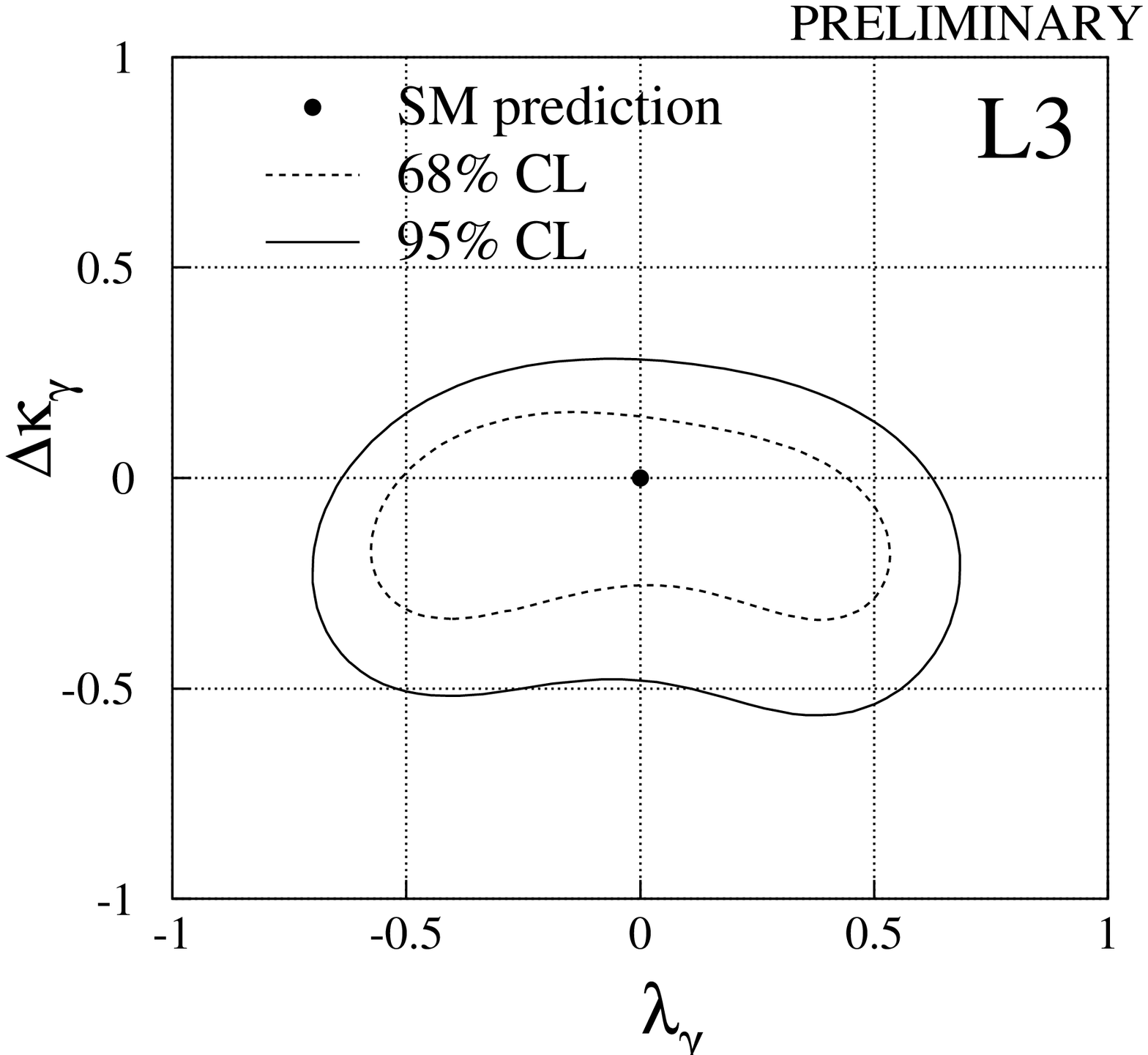,width=0.5\textwidth}} \\
\end{tabular}
}}
\caption{\label{fig:3bis} Evolution of the single W cross section with
  the centre--of--mass energy and determination of the W
  electromagnetic couplings. $\Delta\kappa_\gamma$ stands for the
  deviation of $\kappa_\gamma$ from the SM value of 1.}
\end{figure}

The increasing cross section and the high integrated luminosities
collected at the two energies under investigation allow a good
improvement in the determination of the single W cross section~\cite{singleW2}.
Hadronic decays of the single W are separated from the background
events by means of their kinematic characteristics combined into a
neural network. The single lepton signature is exploited to remove
background events and the distribution of its energy is retained as a
final discriminating variable. A binned maximum likelihood 
fit to these two variables allows to
determine the cross sections for single W production as:
\begin{displaymath}
\sigma_{\rm \epem\rightarrow e\nu_e W}(189\GeV) = 0.53 ^{+0.12}
_{-0.11} \pm 0.03\,{\rm pb},
\end{displaymath}
where the first error is statistical and the second
systematic. Figure~\ref{fig:3bis} compares this cross section and the
lower energy ones to the SM predictions obtained with the EXCALIBUR~\cite{excalibur}
and GRC4F~\cite{grace} MC programs. The experimental precision
approaches the theoretical one. 

With a binned maximum likelihood fit similar to the one used for the
cross section determination, it is possible to measure the
electromagnetic coupling of the W boson as:

\begin{displaymath}
\kappa_\gamma = 0.93 ^{+0.15} _{-0.17};\,\,\,
\lambda_\gamma = -0.30  ^{+0.68} _{- 0.19}.
\end{displaymath}

The precision of $\kappa_\gamma$ is comparable to that of the 
conventional investigation through W pair production. It should be
noted that these measurements come from a two--dimensional fit, whose
contours are also presented in Figure~\ref{fig:3bis}.

%
\section{\bf Two (or more) photons}
%

Multiphoton production in $\epem$ interactions is
dominated by QED also at these high energies. This process  has a
clean experimental signature in the high--performance L3
electromagnetic calorimeter with a negligible background.
The sensitivity of this process to deviations from QED
increases with $\sqrt{s}$, making it well suitable to 
probe New Physics beyond the SM.

The number of observed and expected events with two or more photons in
the  polar angular range $16^\circ < \theta_\gamma < 164^\circ$ and
with energies above $1\GeV$ is reported in Table~3 for the two
energies under study~\cite{gg183189}; perfect agreement is observed
with the SM predictions. From these events the cross sections in the
fiducial volume are extracted as:
\begin{displaymath}
\sigma_{\epem\rightarrow\rm\gamma\gamma(\gamma)}(183\GeV) = 12.2 \pm 0.6\,{\rm pb}
\end{displaymath}
\begin{displaymath}
\sigma_{\epem\rightarrow\rm\gamma\gamma(\gamma)}(189\GeV) = 11.6 \pm 0.3\,{\rm pb}
\end{displaymath}

\begin{table}[h]
  \begin{center}
    \begin{tabular}{|l|c|c||c|c|}
      \hline
      & \multicolumn{2}{c||}{$\sqrt{s} = 183\GeV$} & 
      \multicolumn{2}{c|}{$\sqrt{s} = 189\GeV$}\\
      \hline
      & Observed & Expected  & Observed & Expected \\
      \hline
      2\,$\gamma$ & 436 & 453 & 1302 & 1345 \\
      3\,$\gamma$ & 23 & 24 & 72 & 69 \\
      4\,$\gamma$ & 1 & 0 & 0 & 0 \\
      \hline
    \end{tabular}
    \caption{Number of observed and expected multi--photon events at
      $183\GeV$ and $189\GeV$.}
  \end{center}
\end{table}

\begin{figure}[thb]
\centerline{\mbox{
\begin{tabular}{cc}
\mbox{\psfig{figure=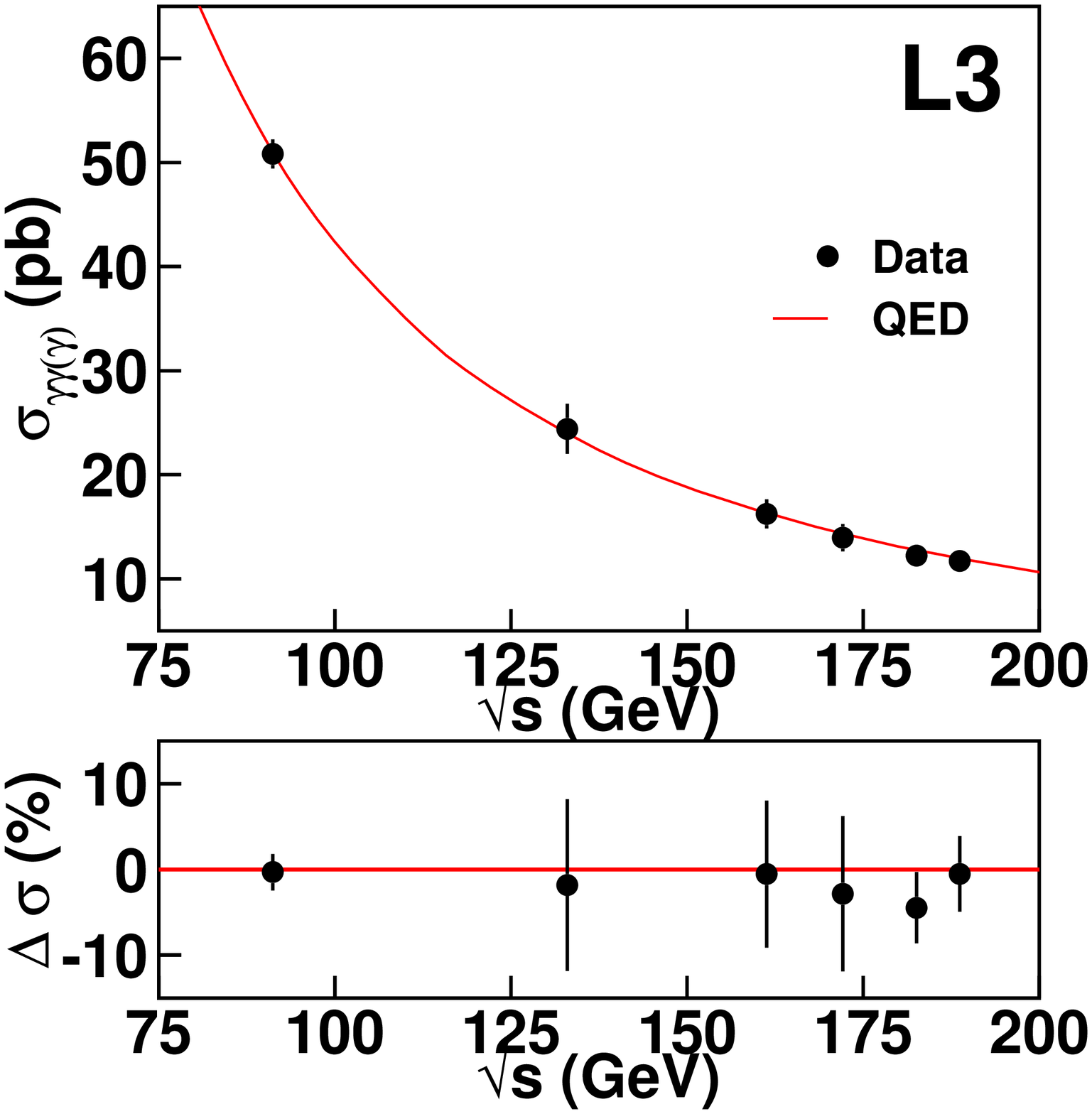,width=0.5\textwidth}} &
\mbox{\psfig{figure=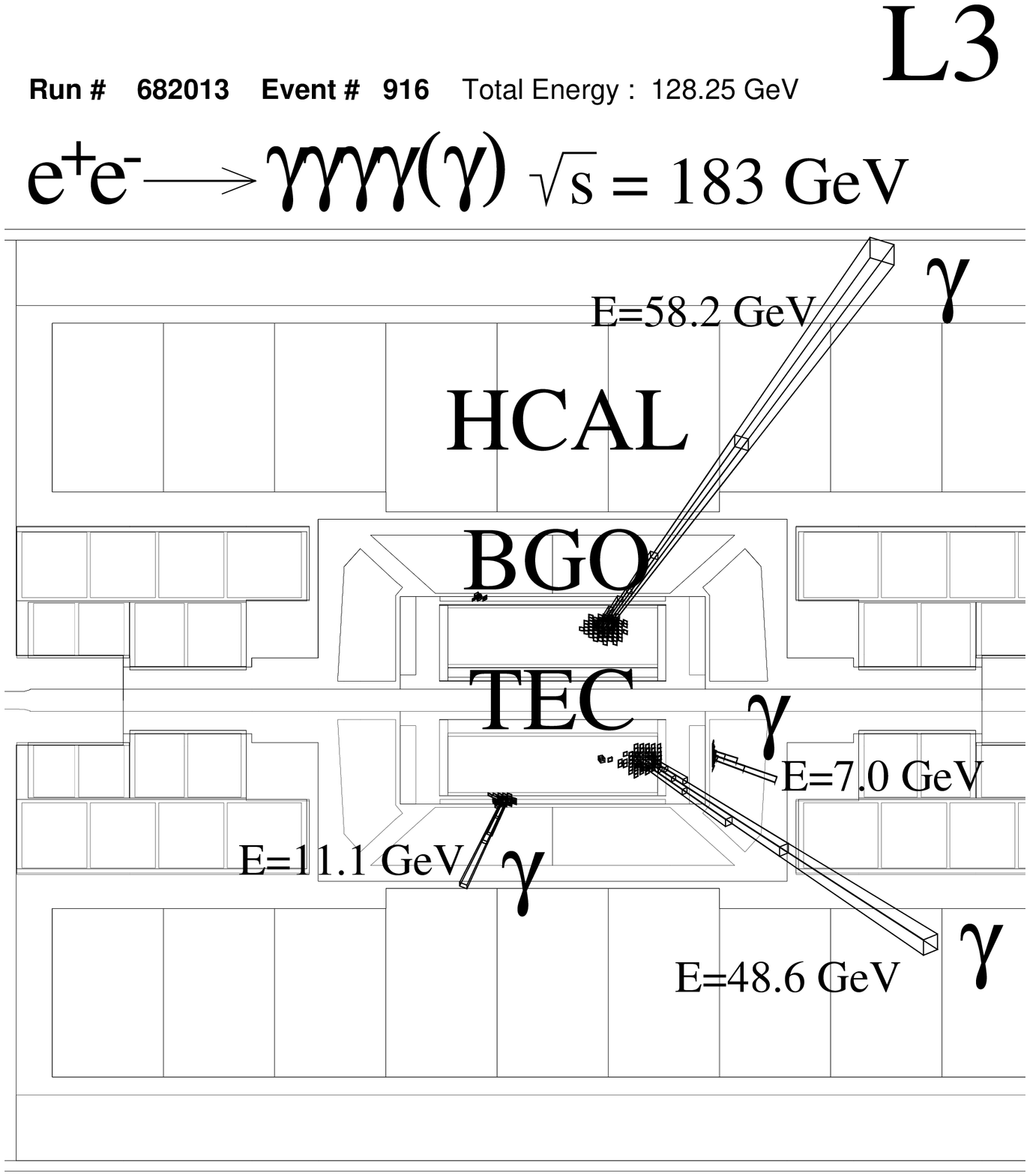,width=0.5\textwidth}} \\
\end{tabular}
}}
\caption{\label{fig:gggg} Evolution of the multiple hard photon
  production cross section with the centre--of-mass energy and side view event display of the
  highest multiplicity selected event.  Four photons are identified, a
  fifth escapes along the beam pipe and a possible sixth photon is detected
with  energy below the identification threshold
set to $1\GeV$ by this analysis.}
\end{figure}

These   cross sections are compared in
Figure~\ref{fig:gggg}a with the QED prediction as a function of
$\sqrt{s}$. Lower energy data~\cite{gglower} are also presented
together with the event display of a selected four
visible photon event. An extra low energy photon is present in the
detector. The kinematics is compatible with a fifth photon escaping
down the uninstrumented beam line.

Deviations from QED are expressed in terms of effective Lagrangians
and affect  the expected differential
cross section ${d \sigma^{\rm QED} / d \Omega}$ by means of an extra
multiplicative term with a different angular structure. Two general forms are
considered~\cite{qeddev}: 
\begin{displaymath}
{d \sigma \over d \Omega} = {d \sigma^{\rm QED} \over d \Omega}
\left(1 + {s^2 \over \alpha}{1 \over  \Lambda^4}(1-\cos^2{\theta_\gamma})
\right)
\end{displaymath}
\begin{displaymath}
{d \sigma \over d \Omega} = {d \sigma^{\rm QED} \over d \Omega}
\left(1 + {s^3 \over 32 \pi \alpha^2}{1 \over \Lambda'^6}
{ 1-\cos^2{\theta_\gamma} \over 1+\cos^2{\theta_\gamma}}
\right),
\end{displaymath}
where $\alpha$ is the electromagnetic coupling and the parameters
$\Lambda$ and $\Lambda'$ have the dimension of an
energy. Alternatively the so called cut--off parameters $\Lambda_{\pm}$ are also 
considered. They are linked to $\Lambda$ by the relation 
$\Lambda^4 = \pm (2/\alpha) \Lambda^4_\pm$. A simultaneous fit to
the differential distribution of the selected events at  the
considered energies and below does not present any deviation from the
SM and the following 95\% CL limits are  extracted on these
deviations from QED:

\begin{displaymath}
\Lambda     > 1304 \GeV;\,\,\,
\Lambda_{+} >  320 \GeV;\,\,\,
\Lambda_{-} >  282 \GeV;\,\,\,
\Lambda^{'} >  702 \GeV.
\end{displaymath}

Photon pairs can be also produced via the $t-$channel exchange of an
excited electron. This interaction is described by a
phenomenological Lagrangian~\cite{ee}  with the scale of the
interaction as a free parameter. Identifying it with the
mass of the excited electron, $m_{\rm e^*}$, limits at 95\% CL  are
set as:

\begin{center}
\begin{tabular}{cc}
$m_{\rm e^*} > 323\GeV$ (non-chiral) &
$m_{\rm e^*} > 282\GeV$ (chiral), \\
\end{tabular}
\end{center}
according to the chirality of the interaction.

%
\section{\bf One photon and one Z}
%

The non Abelian structure of the SM predicts the existence at tree
level of the
triple vertex of one neutral and two charged bosons, $\gamma$WW and
ZWW, whose presence is successfully verified at LEP, as already reported
above. The triple vertices of neutral bosons, $\gamma\gamma$Z,
$\gamma$ZZ, and ZZZ are
on the other hand forbidden at tree level in the SM. The possible
presence of the first two of them, is probed by
the associated production of a Z boson and a  photon in $\epem$
collision. In the SM this process takes place via
the $t-$channel electron exchange, that is nothing else than  the
radiative return to the Z process described above, yielding almost
monoenergetic photons. The SM cross section
of this process decreases with $\sqrt{s}$ while the possible
anomalous contribution coming from a $t-$ channel Z or photon exchange
does not, making the study of this process of
interest with the increase of the LEP beam energy.

The  Z$\gamma$V vertex, with $V$ either a photon or a Z, is
parametrised as~\cite{hagiwara,gounaris}: 

\begin{displaymath}
\Gamma^{\alpha\beta\mu}_{\Zo\gamma V}(q_1,q_2,P) = i {{s - m_V^2} \over {m^2_\Zo}} \times
\end{displaymath}
\begin{displaymath}
\big\{  h_1^V ( q_2^\mu g^{\alpha\beta} - q_2^\alpha g^{\mu\beta}) + { 
h_2^V  \over m^2_\Zo} P^\alpha  (P\cdot q_2 \,g^{\mu\beta}- q_2^\mu P^{\beta})
 +   h_3^V  \epsilon^{\mu\alpha\beta\rho} q_{2\rho}  
- { h_4^V  \over m^2_\Zo}  P^\alpha \epsilon^{\mu\beta\rho\sigma} P_\rho q_{2\sigma} \big\},
\end{displaymath}
where $q_1$, $q_2$ and $P$ are the four--momenta of the  Z, $\gamma$ and
$V$ bosons. Eight couplings appear in the above expression, four
for each $V$ boson. As in the W case, only the real part of these
couplings are of 
interest in absence of indications for New Physics. The couplings $ h_1^V $ and $ h_2^V $ are zero in
the SM at tree level and violate the CP symmetry while  $ h_3^V $ and $ h_4^V $
preserve it and  have an estimated value of
$10^{-4}$  from higher order SM processes. The effect of a value of these couplings different from
zero would manifest itself as an enhancement of the number of Z$\gamma$
events, more pronounced for photons emitted at a large polar
angle. 

The experimental selection of Z$\gamma$ events~\cite{zg} 
proceeds in the highest branching ratio channels $\rm q\bar{q}\gamma$
and $\nu\bar{\nu}\gamma$. A good agreement is found between the number of
selected events and the SM expectations. 

Five variables describe the phase space of the $\rm f \bar{f} \gamma$
events, the energy and the two angles of the photon and the two angles
of one of the fermions in the Z reference frame. An unbinned maximum
likelihood fit is performed in this five dimensional space for each of the eight couplings, fixing
the other seven to zero. All the
measured values are consistent with the SM and 95\% CL limits are
extracted as:

\begin{displaymath}
-0.16  \leq h_1^{\Zo}    \leq 0.09;\   \ 
-0.07  \leq h_2^{\Zo}    \leq 0.11;\   \
-0.22  \leq h_3^{\Zo}    \leq 0.10;\   \ 
-0.06  \leq h_4^{\Zo}    \leq 0.14;
\end{displaymath}
\begin{displaymath}
-0.12  \leq h_1^{\gamma}    \leq 0.09;\   \ 
-0.06  \leq h_2^{\gamma}    \leq 0.07;\   \
-0.12  \leq h_3^{\gamma}    \leq 0.01;\   \ 
-0.01  \leq h_4^{\gamma}    \leq 0.09.
\end{displaymath}

\begin{figure}[t]
\centerline{\mbox{
\begin{tabular}{cc}
\mbox{\psfig{figure=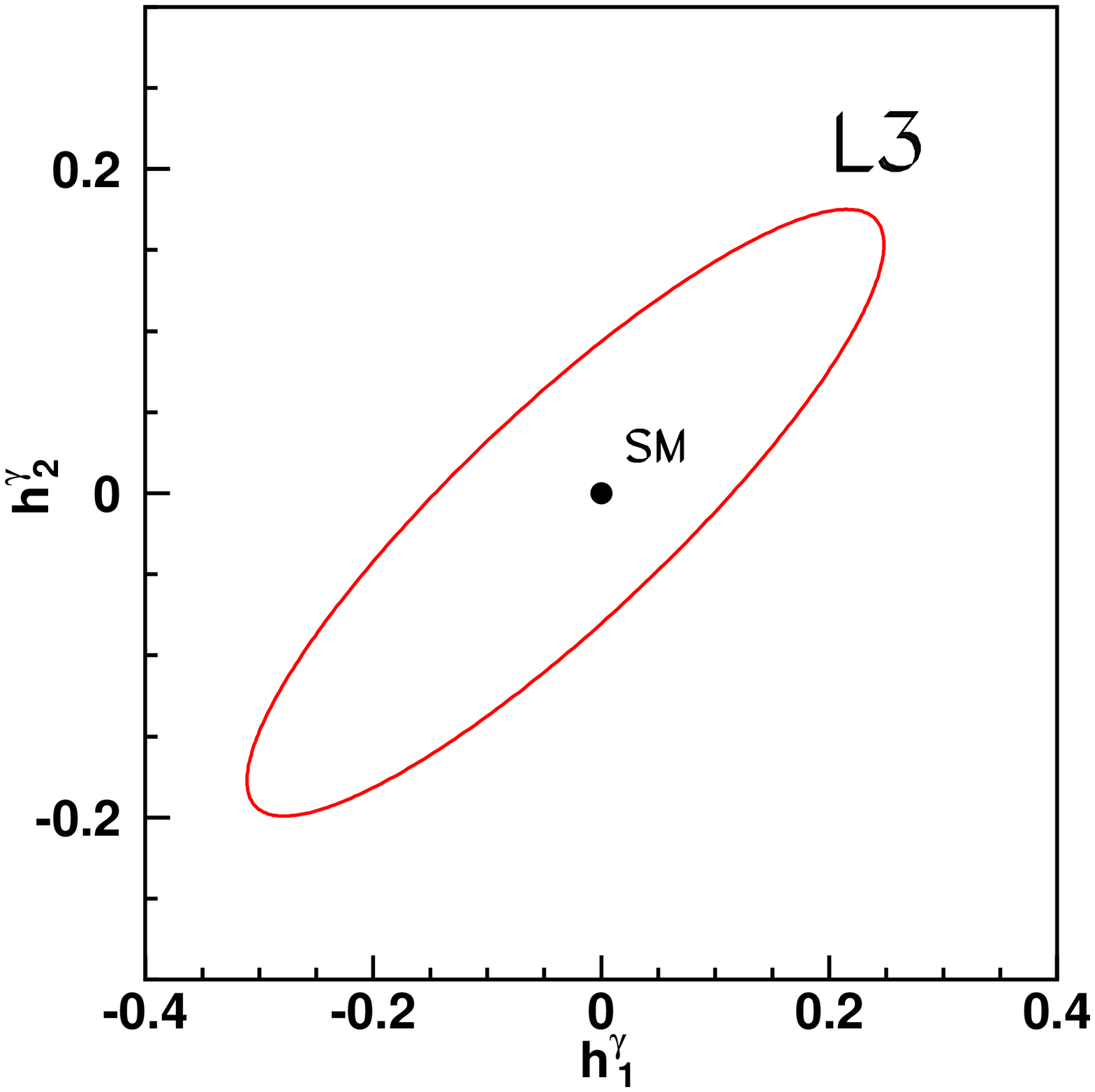,width=0.4\textwidth}} &
\mbox{\psfig{figure=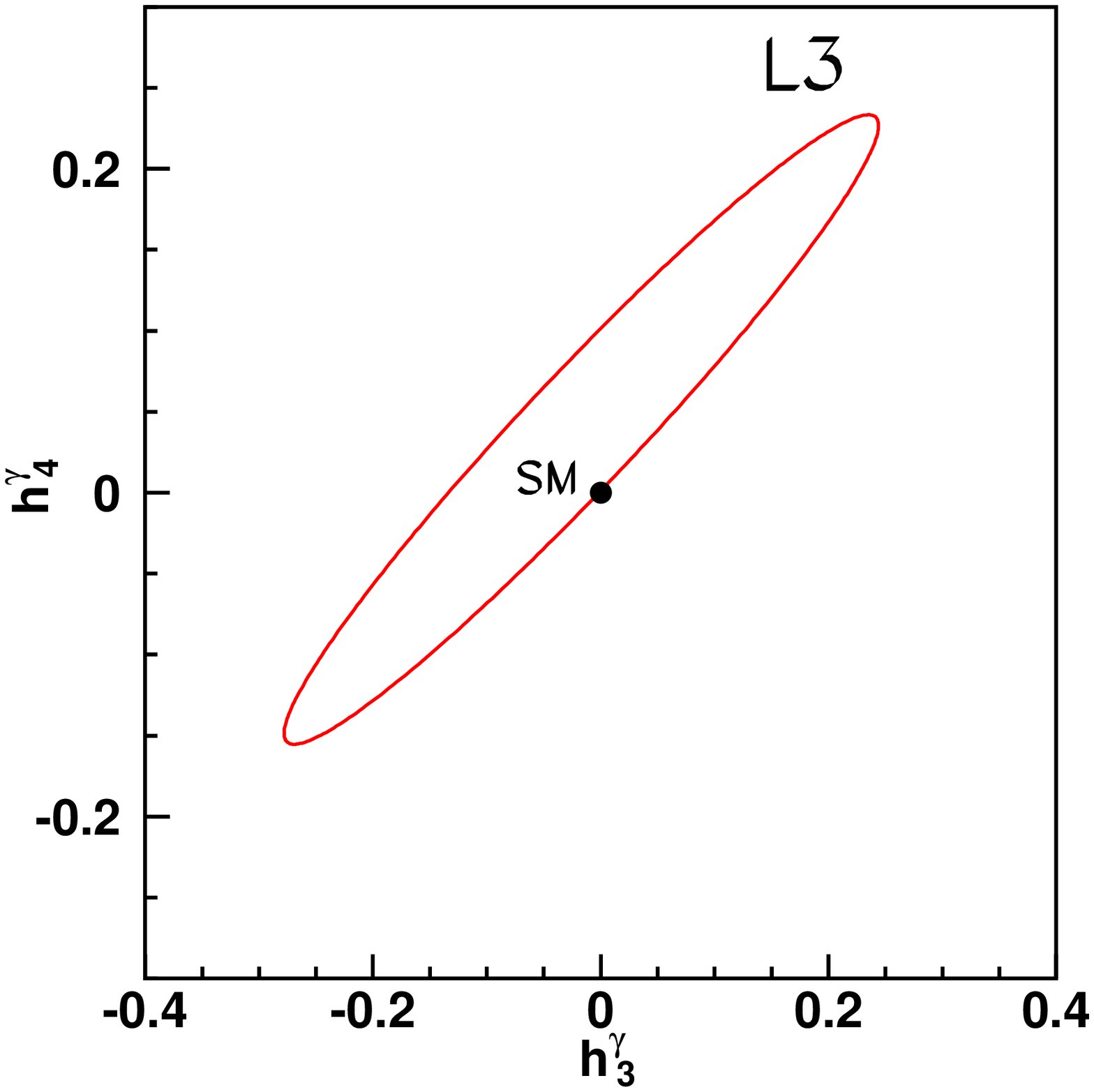,width=0.4\textwidth}} \\
\mbox{\psfig{figure=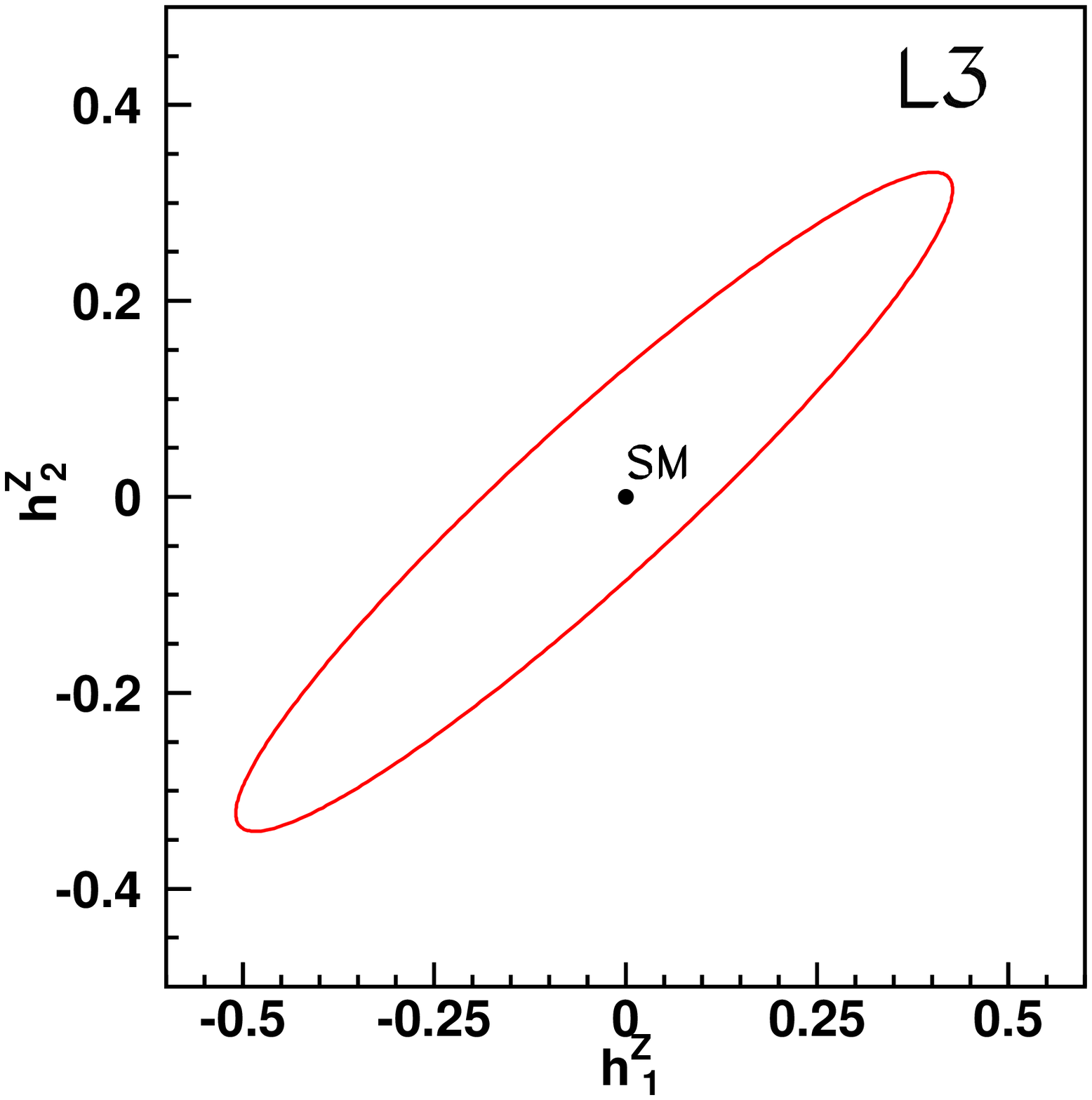,width=0.4\textwidth}} &
\mbox{\psfig{figure=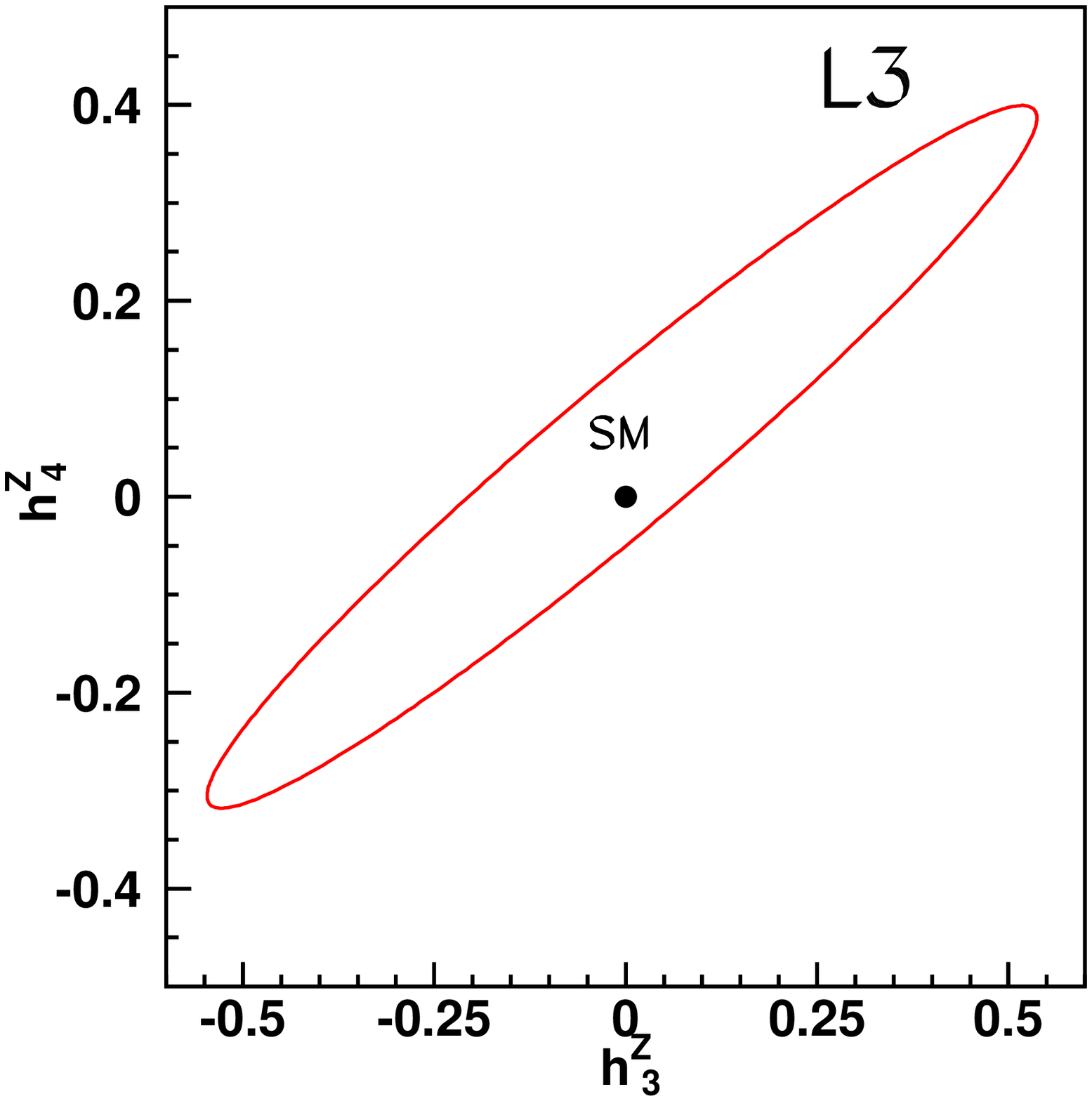,width=0.4\textwidth}} \\
\end{tabular}
}}
\caption{\label{fig:zg} 95\% CL contours from simultaneous
  two--parameter fits to the coupling with the same CP--parity
  involving the same exchanged vector boson.} 
\end{figure}

Figure~\ref{fig:zg} presents the 95\% CL contours if a two--parameter
fit is performed on the pairs of couplings with the same CP--parity
and involving the exchange of the same neutral boson.

%
\section{\bf Two Z}
%

The data set under investigation was collected above  the production threshold
of Z boson pairs. This process is 
of particular interest 
as it constitutes an irreducible background for the search of the
SM Higgs boson and to several other processes predicted by
theories beyond the SM.  In addition it allows the
investigation of possible triple neutral gauge boson couplings, ZZZ and
ZZ$\gamma$, forbidden by the SM.

The experimental investigation of ZZ production~\cite{l3zz183,l3zz189}
is made difficult by
its rather low cross section, compared with competing processes
that constitute large and sometimes irreducible
backgrounds.

The  Z pair signal is defined starting from 
generator level 
phase--space cuts on four fermion final states: the invariant mass of both
generated fermion pairs must be between $70\GeV$ and
$105\GeV$. This criterion has to be satisfied by at least one of the two
possible pairings of four same flavour fermions. In the case in which
fermion pairs can originate from a charged--current process 
the masses of the fermion pairs susceptible to
come from W decays  are required to be either below  $75\GeV$
or above $85\GeV$. Events with electrons in the final state are
rejected if $|\cos{\theta_{\rm e}}| > 0.95$, where $\theta_{\rm e}$
is the electron polar angle.

The expected cross sections for the different final states are
computed  with EXCALIBUR MC  and amount
to a total of 0.25\,pb at $183\GeV$  and 0.66\,pb at $189\GeV$.

\begin{figure}[H]
\centerline{\mbox{
\begin{tabular}{cc}
\mbox{\psfig{figure=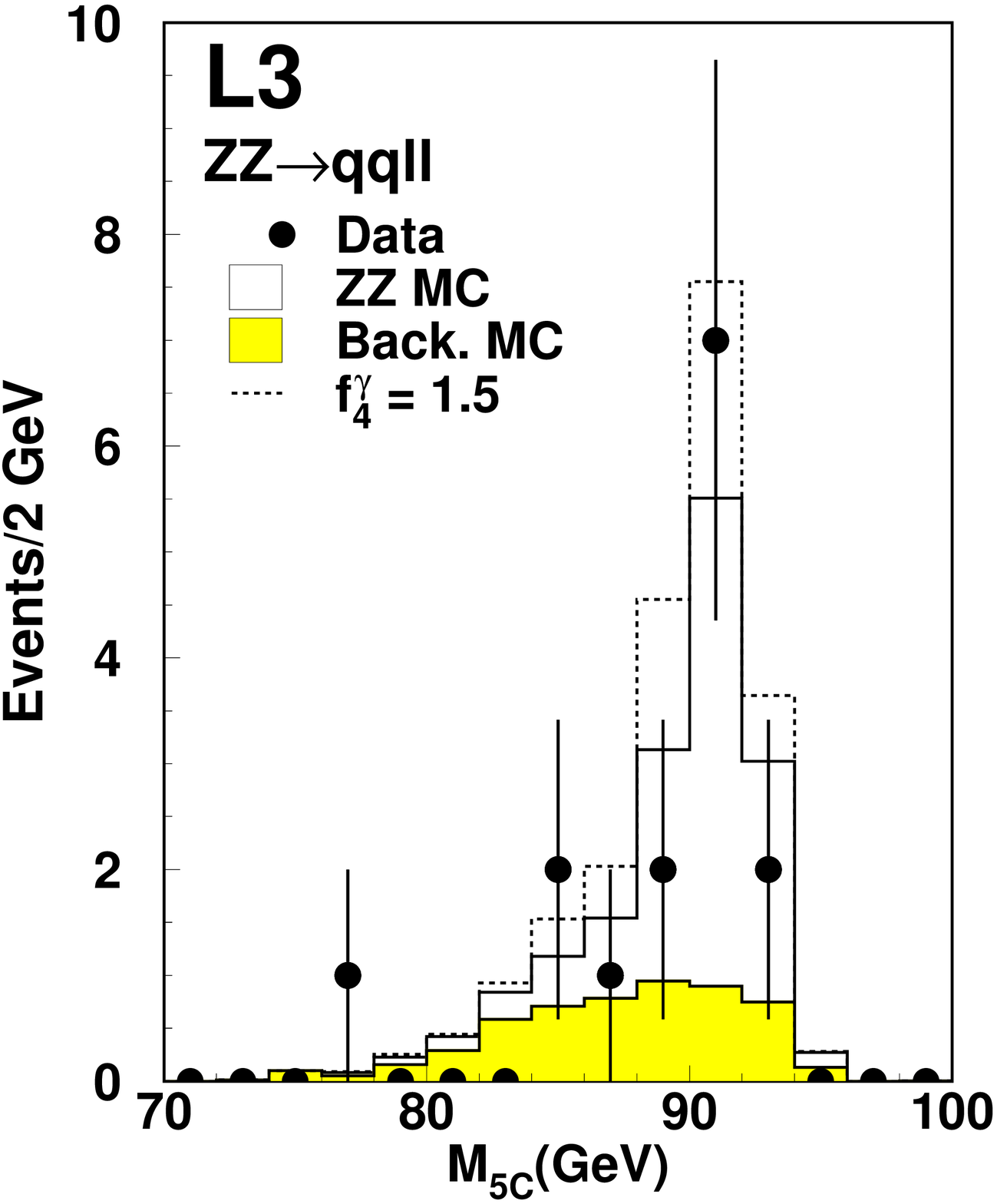,width=0.4\textwidth}} &
\mbox{\psfig{figure=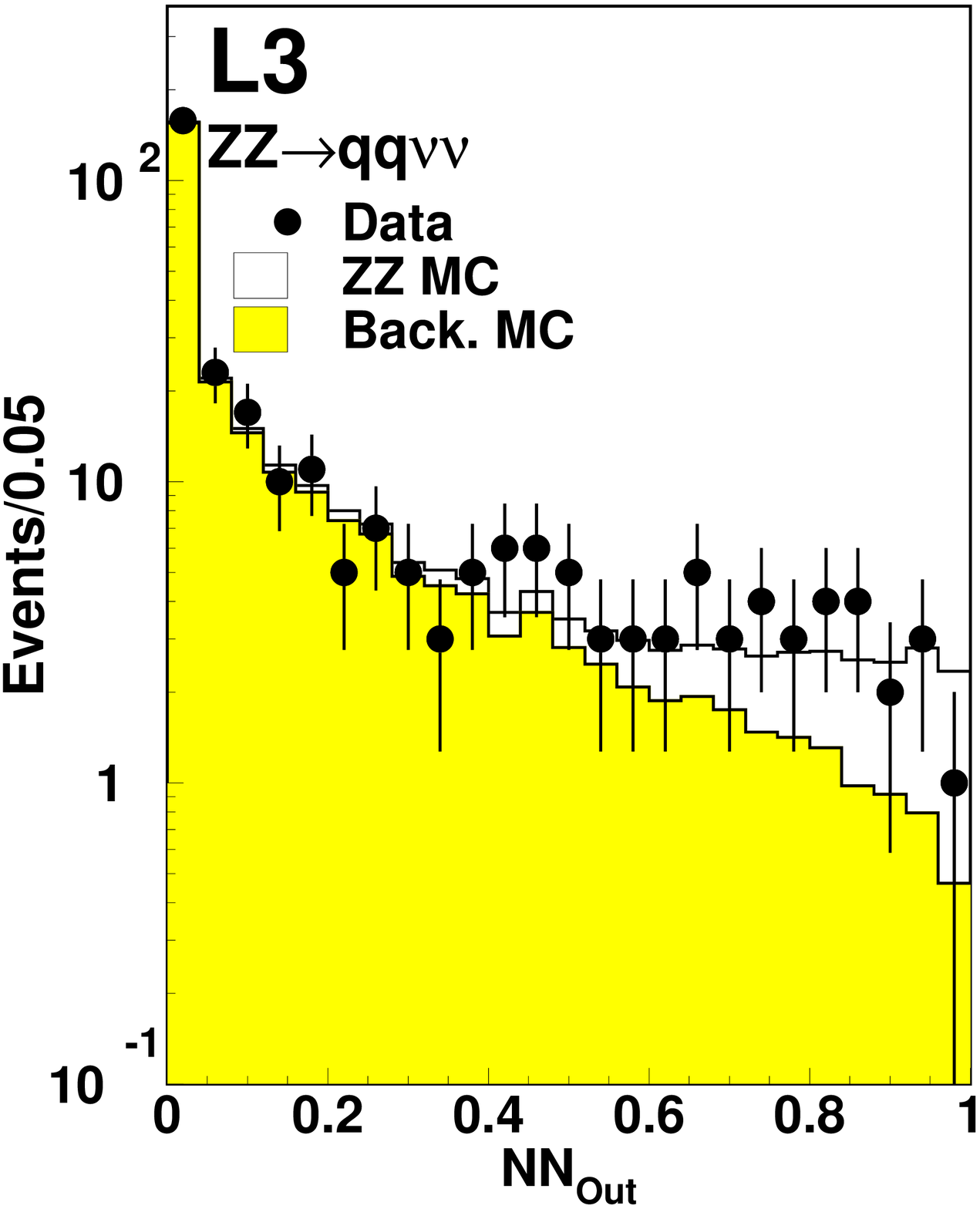,width=0.4\textwidth}} \\
\mbox{\psfig{figure=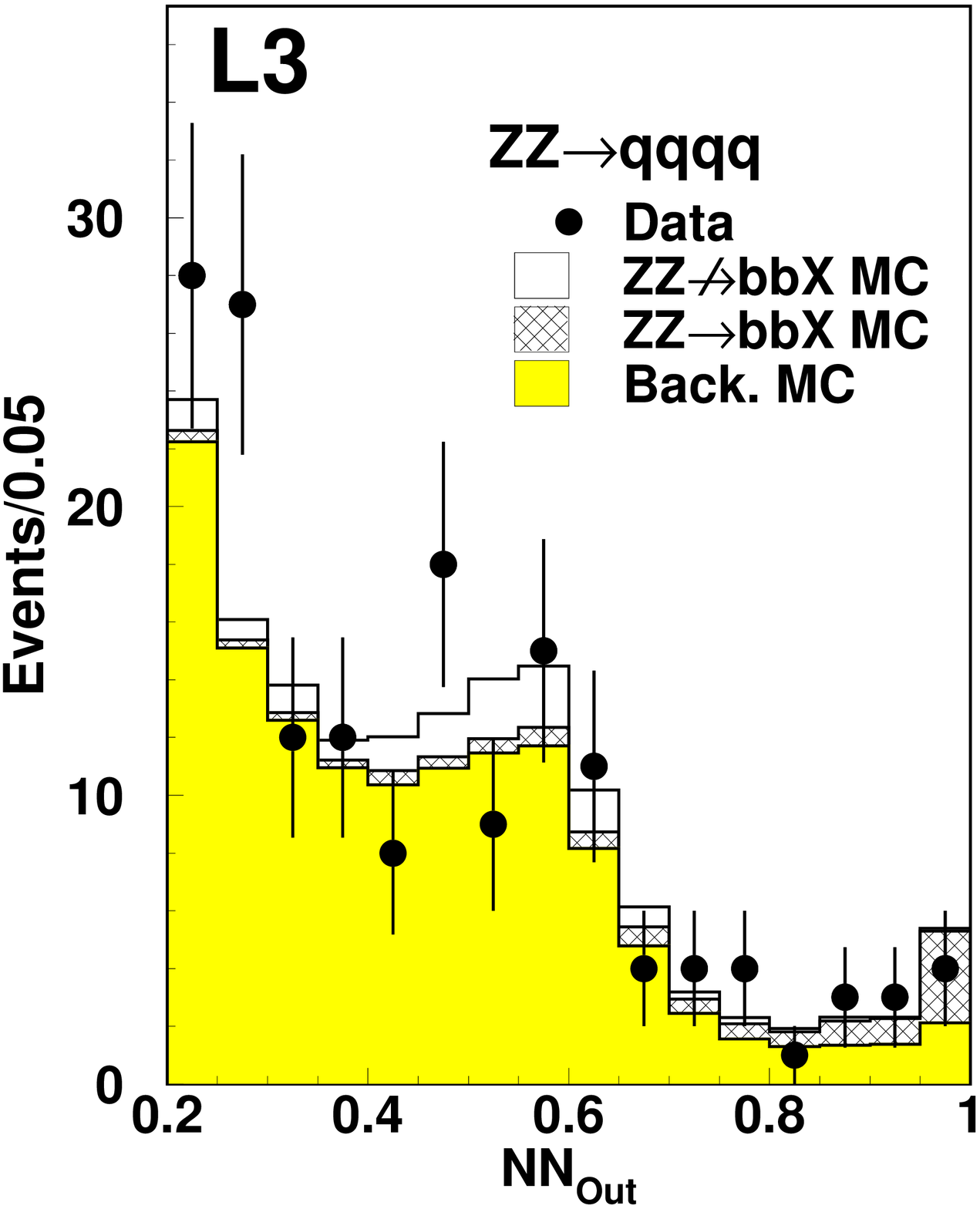,width=0.4\textwidth}} &
\mbox{\psfig{figure=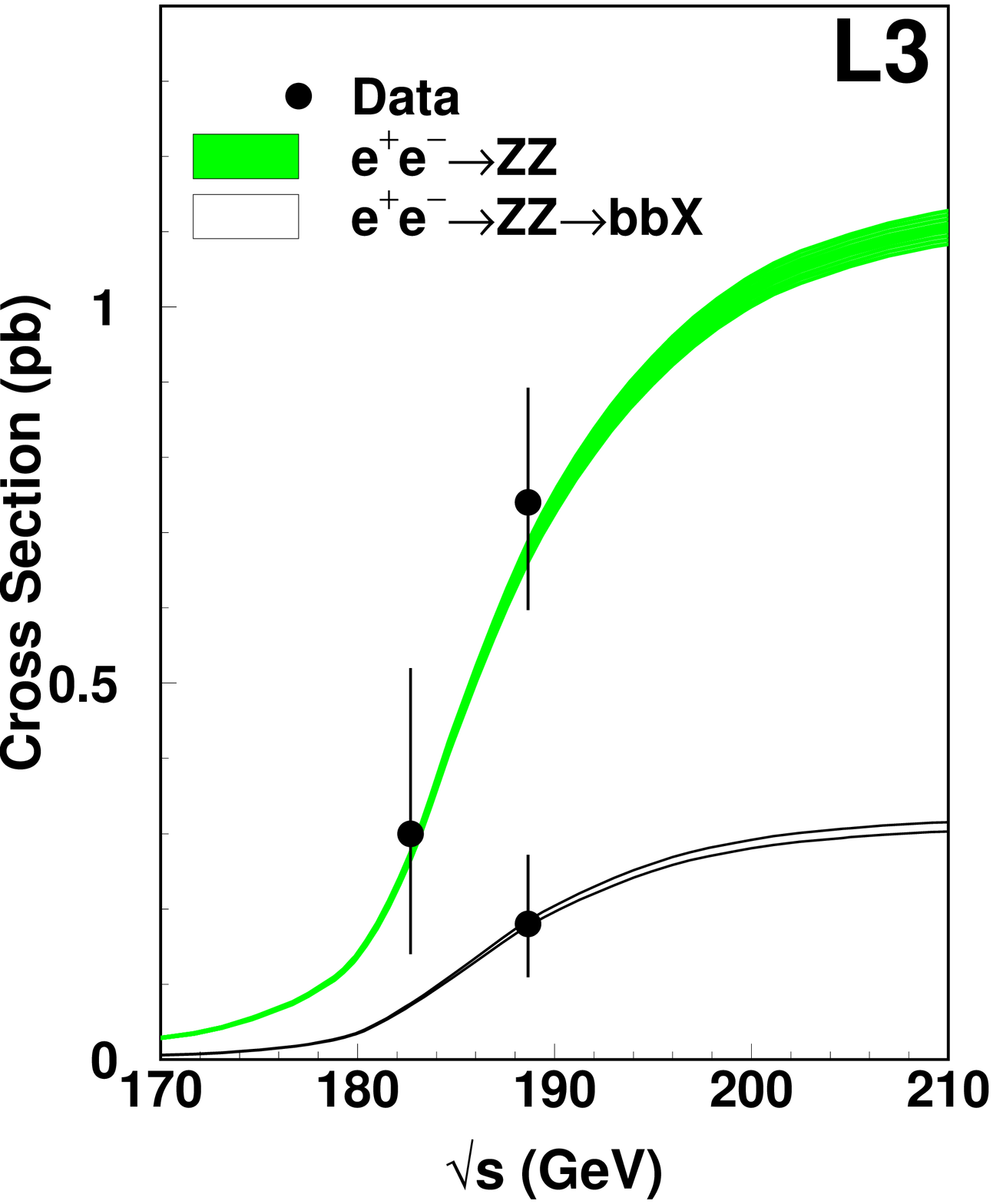,width=0.4\textwidth}} \\
\end{tabular}
}}
\caption{\label{fig:3} Fit Z mass in $\eeZZto\qqll$ events in data and
  MC together with the distortion expected from anomalous couplings
  contributions, neural
  network outputs for the selection of $\eeZZto\qqnn$ and
  $\eeZZto\qqqq$ events and evolution of the $\epem\rightarrow\Zo\Zo$
  and 
  $\epem\rightarrow\Zo\Zo\rightarrow\rm b \bar{b} X$ cross sections with the
  centre--of--mass energy.}
\end{figure}

All the visible final states of  Z pair decay are
investigated. These selections are based on the identification of two fermion pairs,
each with a mass close to the Z boson mass, and are different at the
two centre--of--mass energies to account for  the different signal topology due to the
larger boost of the Z bosons at $189\GeV$. This boost leads to acollinear and acoplanar
fermion pairs. Kinematic fits help in checking the hypothesis of equal
mass particles. The $\qqnn$ and $\qqqq$ final states are selected by
combining the kinematic variables into a neural network.
Figure~\ref{fig:3} presents the reconstructed mass $M_{5C}$ of
selected ZZ$\rightarrow\qqll$ events at $189\GeV$ after the kinematic
fit and the output of the two neural networks for the  $\qqnn$ and
$\qqqq$ selections.

\begin{figure}[thb]
\centerline{\mbox{
\begin{tabular}{cc}
\mbox{\psfig{figure=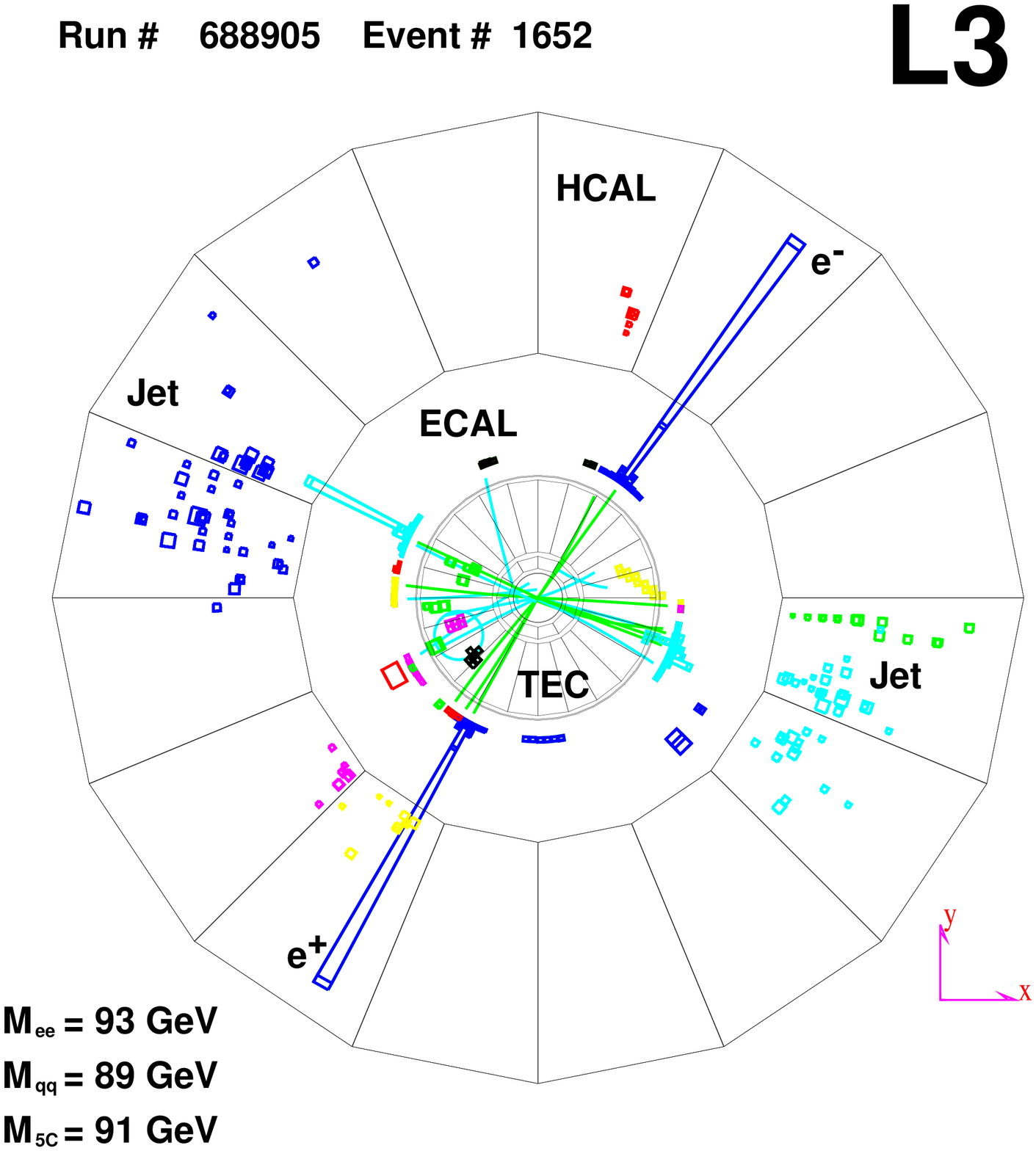,width=0.5\textwidth}} &
\mbox{\psfig{figure=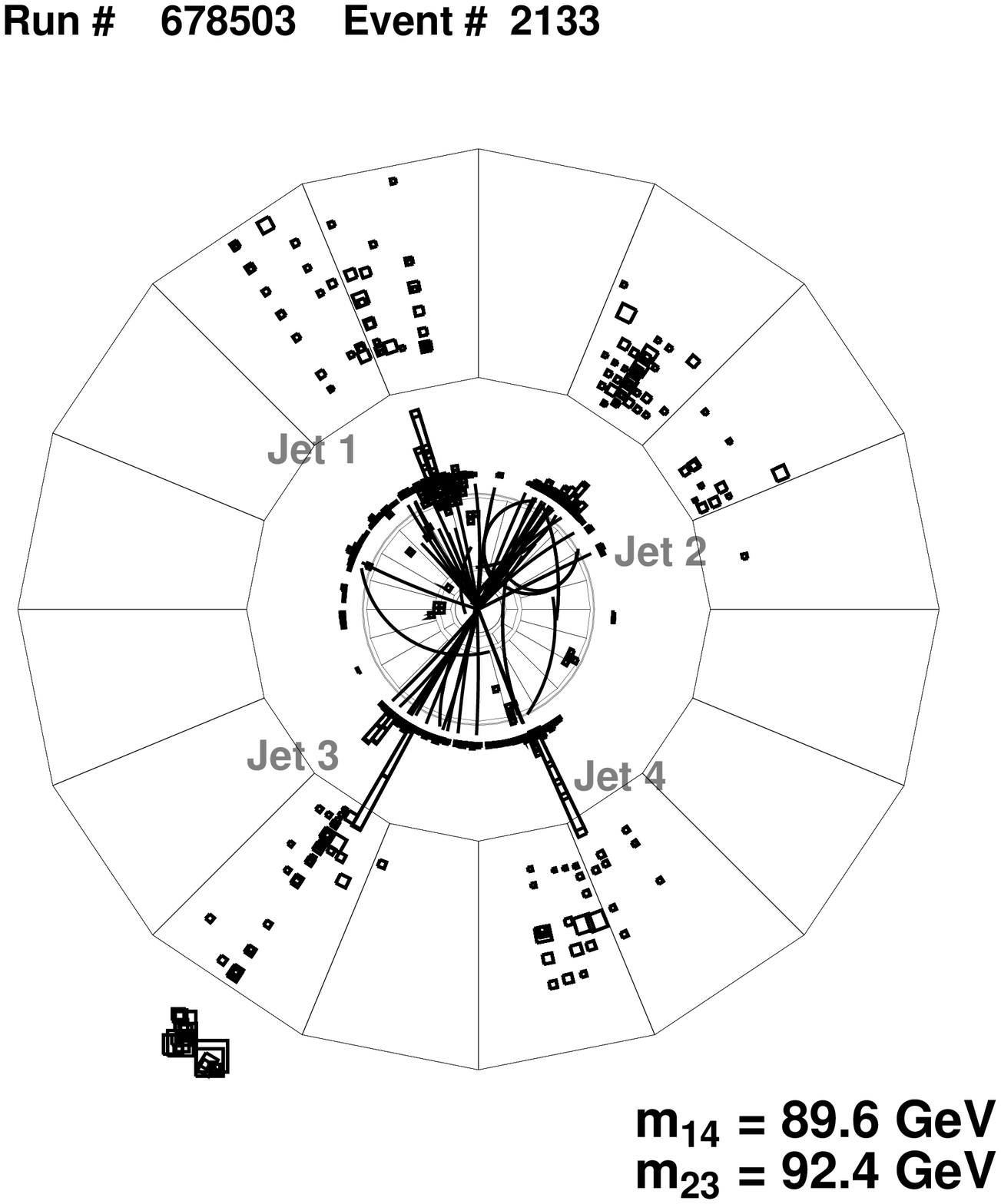,width=0.5\textwidth}} \\
\end{tabular}
}}
\caption{\label{fig:zz} Data events selected by the  $\eeZZto\qqll$
  and  $\eeZZto\qqqq$.}
\end{figure}

The ZZ cross sections are determined by a binned maximum likelihood
fit to the most discriminating variables of each selection. Within the
above mentioned signal definition cuts the results are:
\begin{displaymath}
\sigma_{\epem\rightarrow\rm ZZ}(183\GeV) = 0.30
^{+0.22\,\,+0.07}_{-0.16\,\,-0.03}\,{\rm pb}
\end{displaymath}
\begin{displaymath}
\sigma_{\epem\rightarrow\rm ZZ}(189\GeV) = 0.74 ^{+0.15}_{-0.14} \pm 0.04\,{\rm pb},
\end{displaymath}
where the first errors are statistical and the second systematic. The
first of these values constitutes the first observation of the
$\epem\rightarrow\rm ZZ$ process. Figure~\ref{fig:zz} presents two of
the selected data events at $\sqrt{s} = 183 \GeV$.

It is of particular interest to investigate the rate 
of ZZ events with  b
quark content. The production of the minimal or a supersymmetric Higgs
boson
would give an enhancement of these events and their
study on one hand complements the dedicated search for such
processes~\cite{h189,ha189} and on the other hand proves the
experimental sensitivity to such a signal. The expected Standard Model cross
section for the ZZ$\rm \rightarrow b\bar{b}X$  final states at
$189\GeV$ is  0.18\,pb.  

The investigation of the ZZ$\rm \rightarrow b\bar{b}X$ events proceeds by
complementing the analyses of the $\qqnn$ and $\qqll$ final states
with a further variable describing the b quark content in
the event~\cite{h189,ha189}, while the $\qqqq$ selection already
includes such information to partially reject the W pair background.
The combination of these selections yields:

\begin{displaymath}
\sigma_{\rm ZZ\rightarrow b\bar{b}X}(189\GeV)=0.18^{+0.09}_{-0.07}\pm 0.02\,\mathrm{pb}.
\end{displaymath}
The first error is statistical and the second systematic. This result agrees with the SM expectation and differs
from zero at 99.9\% confidence level.
Figure~\ref{fig:3} displays these  cross
sections and their expected 
evolution with $\sqrt{s}$.

A parametrisation of the  ZZZ and ZZ$\gamma$ anomalous couplings is
given in References~\cite{hagiwara,gounaris}. 
Assuming on-shell production of a pair of Z bosons, only four couplings
$f_i^{\mathrm V}~(i=4,5 ; {\mathrm V}=\gamma,\Zo)$, 
where the V superscript corresponds to an anomalous coupling $\rm ZZV$, 
may be different from zero. At tree level these couplings are zero in
the SM. As in the cases already described above only the real part of
these coupling is of interest, if no deviations from the SM are
observed.
The $f_5^{\mathrm V}$ couplings violate the C-- and
P--symmetries, preserving the CP-- one that is instead violated by the 
$f_4^{\mathrm V}$ couplings. These couplings are independent from the
$h_i^\Zo$ ones that parametrise the possible anomalous ZZ$\gamma$ vertex~\cite{hagiwara,gounaris},
whose investigation was described above.

In order to calculate the impact of anomalous couplings
on the measured distributions in 
the process
$\rm e^+e^- \rightarrow f \bar{f} f' \bar{f}'$, 
the EXCALIBUR generator is extended~\cite{madridpaper} and used to
reweight the SM MC events.
Figure~\ref{fig:3} displays the effects of an
anomalous value of $f_4^{\gamma}$ obtained by reweighting with this
technique the four--fermion MC events selected by the $\qqll$ analysis.
 
The  anomalous couplings not only change
the ZZ  cross section  but also the shape of the
distributions of several kinematic variables describing the process. A
binned maximum likelihood fit is performed on the same discriminating
distributions used to determine the cross sections at $183\GeV$ and
$189\GeV$. A coupling   $f_i^{\mathrm V}$ is left free in the fit, fixing the
others to zero. The results of these fits are
compatible with the SM values and 95\% CL limits on the
couplings are set as:
\begin{displaymath}
-1.9  \leq f_4^{\Zo}    \leq 1.9;\   \ 
-5.0  \leq f_5^{\Zo}    \leq 4.5;\   \
-1.1  \leq f_4^{\gamma} \leq 1.2;\   \ 
-3.0  \leq f_5^{\gamma} \leq 2.9.
\end{displaymath}
These limits are still valid for off--shell ZZ production where
additional couplings are possible. 
The small asymmetries in these limits are due to the interference
term between the anomalous coupling diagram and the Standard 
Model diagrams.

%
\section{\bf ...in the Extra Dimensions}
%

Several of the results described above are interpreted in the
framework of recent theories of the gravitational interaction that
introduce extra spatial dimensions. The scene for them is set by two
observations. The first is the huge difference between the scales of
two of the fundamental interactions of nature, the electroweak ($M_{ew}\sim
10^{2}\GeV$) typical of the SM and the Planck scale ($M_{Pl}\sim
10^{19}\GeV$) linked to the gravitational constant. The second observation is that
collider experiments have successfully tested the 
SM at its
characteristic distance $M_{ew}^{-1}$ while 
the experimental study of the gravitational
force extends only down to distances  of the order of a
centimetre~\cite{expgravity} thirty
three orders of magnitude  above the  distance
$M_{Pl}^{-1}$. 

If  $n$ extra spatial dimensions of
size $R$ are postulated, the scale of gravity,  $M_S$, may be assumed
to be of the same order as $M_{ew}$  explaining the observed 
difference between $M_{ew}$ and $M_{Pl}$~\cite{arkani}. 
In this Low
Scale Gravity (LSG) scenario   the macroscopic expectations of
gravity are preserved
by the application of the Gauss' theorem in the extra dimensions:
\begin{equation}
M_{Pl}^2 \sim R^n M_S^{n+2}.
\end{equation}

\begin{figure}[H]
\centerline{\mbox{
\begin{tabular}{cc}
\mbox{\psfig{figure=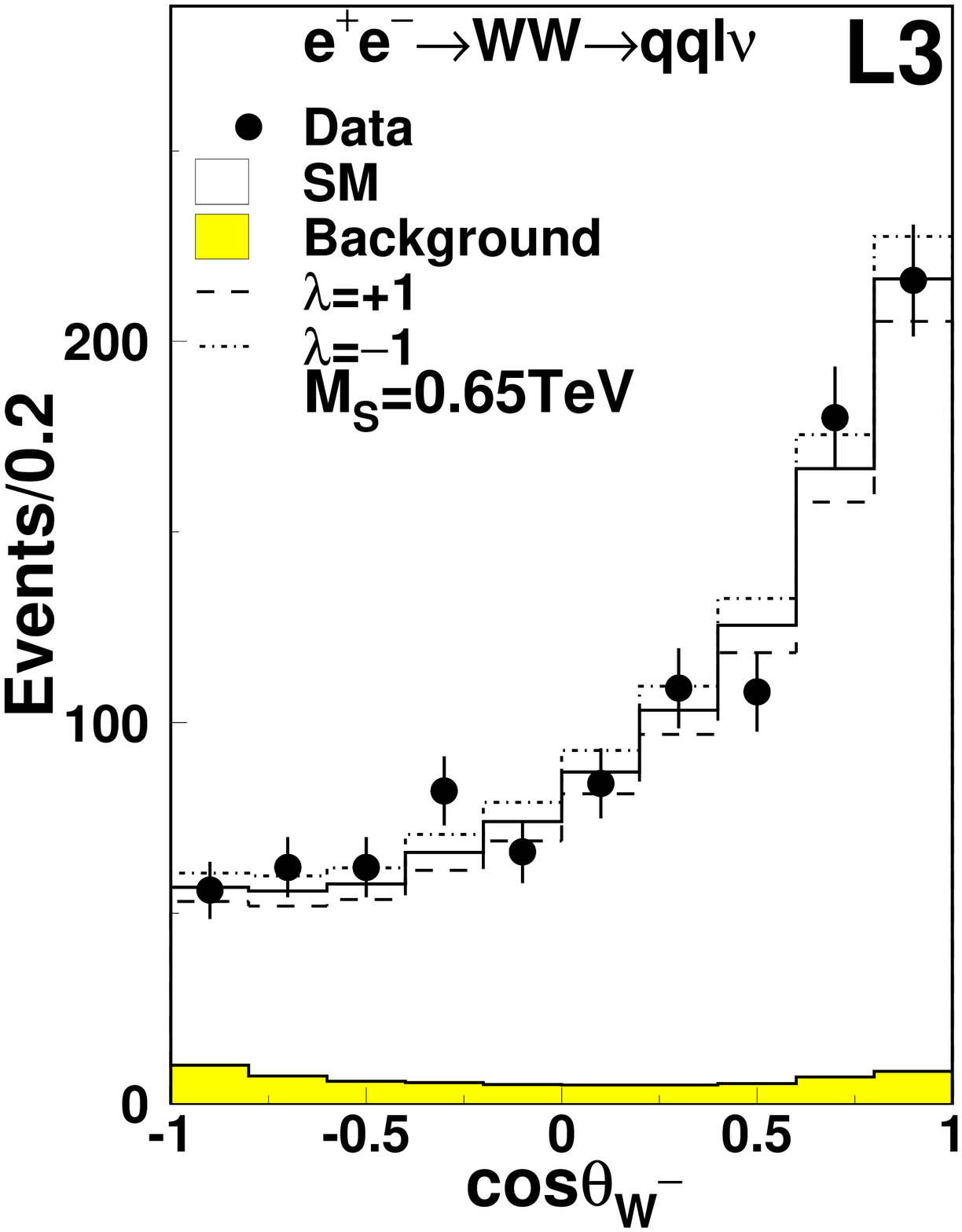,width=0.4\textwidth}} &
\mbox{\psfig{figure=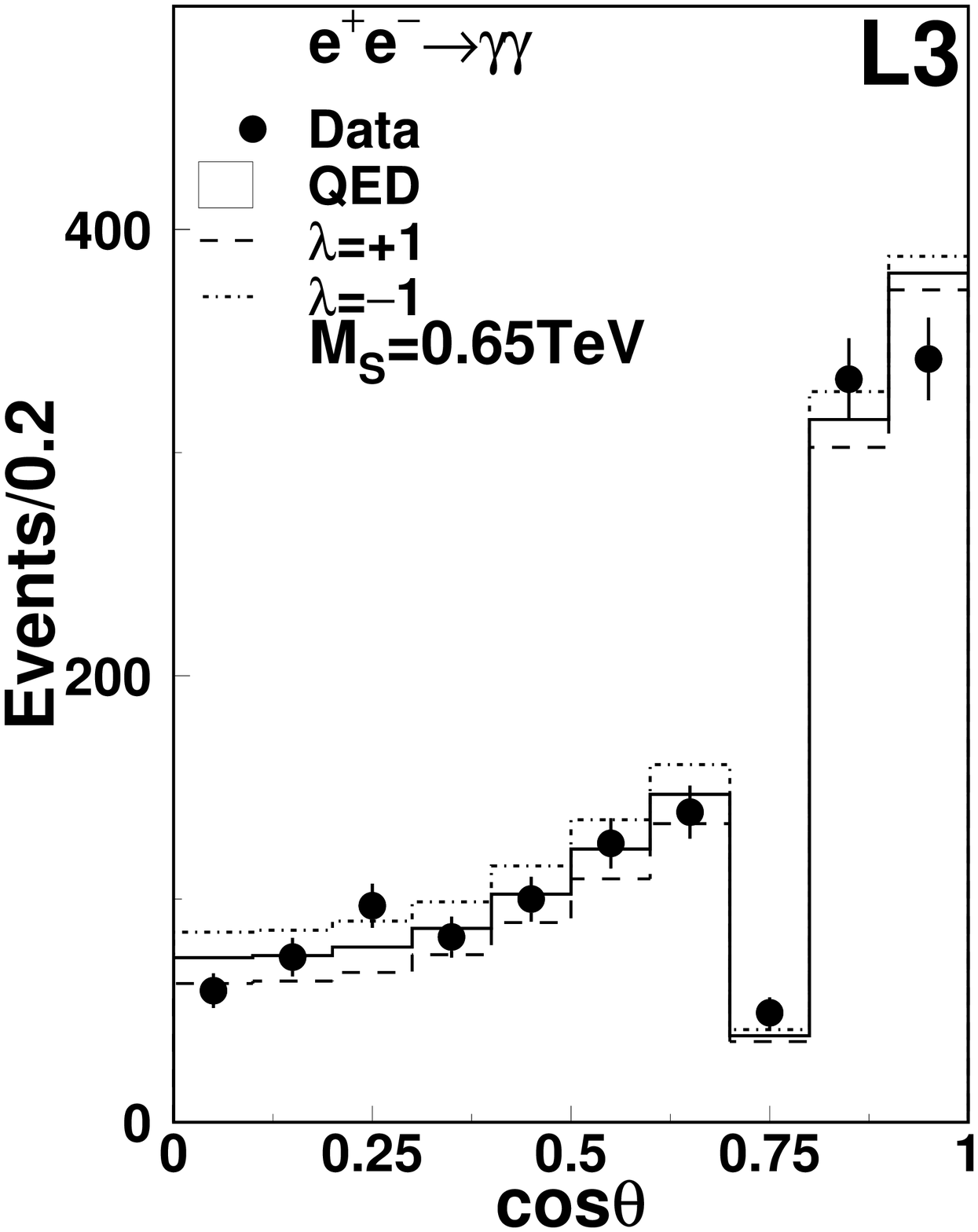,width=0.4\textwidth}} \\
\mbox{\psfig{figure=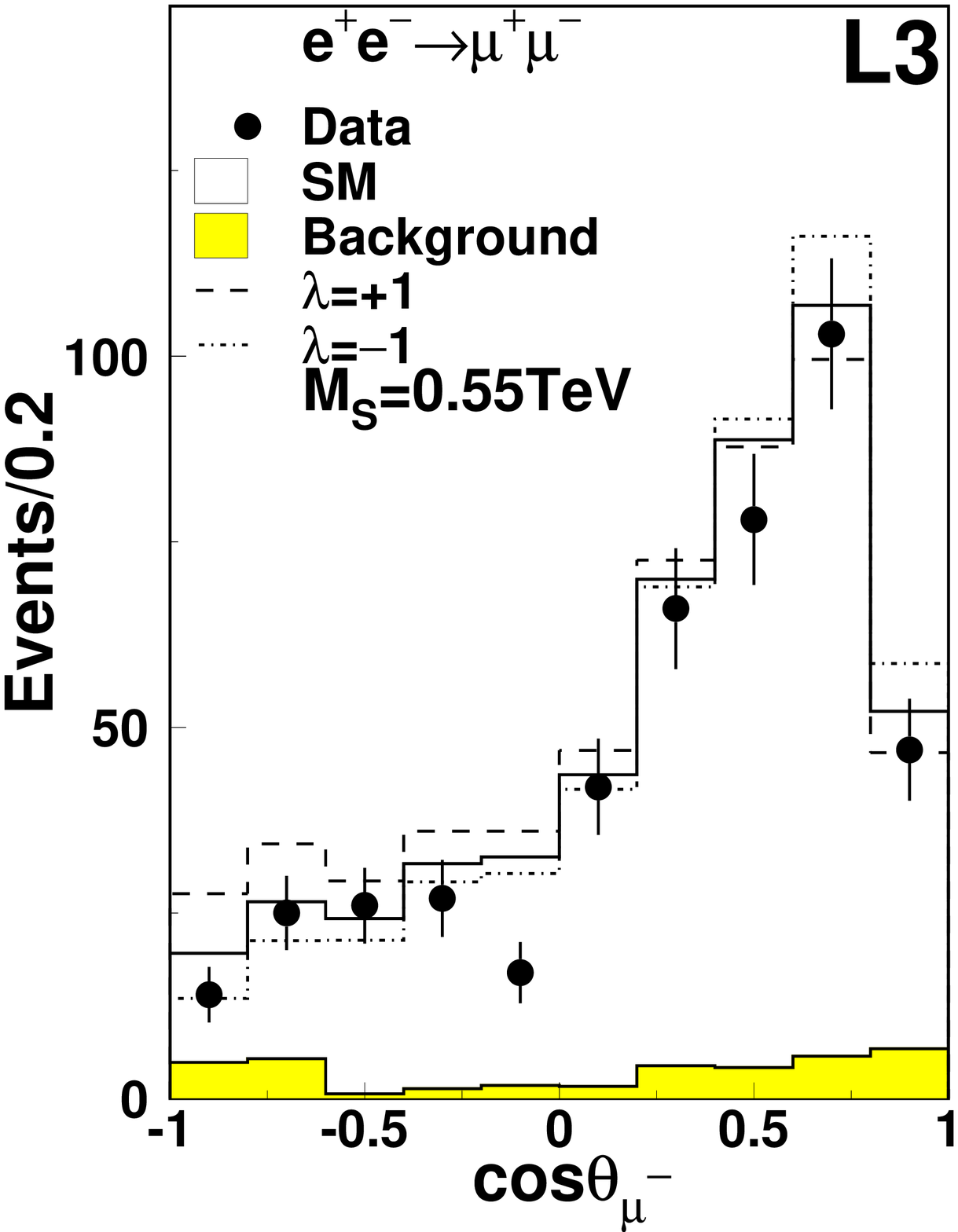,width=0.4\textwidth}} &
\mbox{\psfig{figure=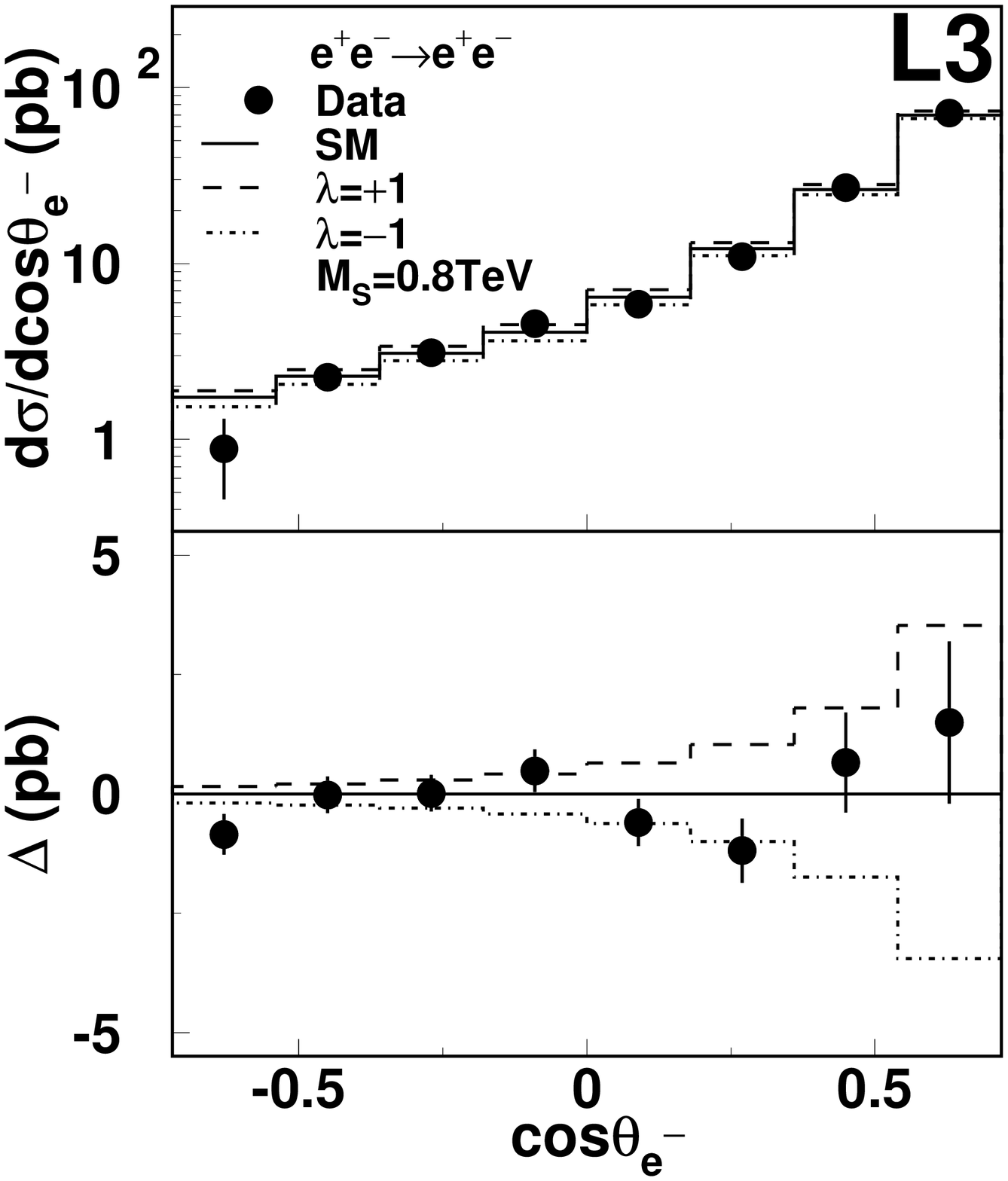,width=0.4\textwidth}} \\
\end{tabular}
}}
\caption{\label{fig:5} Differential distributions of
  $\epem\rightarrow\rm WW\rightarrow q\bar{q}' \ell \nu$,
  $\epem\rightarrow\gamma\gamma$ and 
  $\epem\rightarrow\mu^+\mu^-$ events and differential cross section
  for the $\epem\rightarrow\epem$ final
  state.  SM
  expectations and LSG distortions are plotted, for both signs of
  their interference with SM processes. Data at $189\GeV$ data are shown.}
\end{figure}

The LSG scenario predicts the size $R$ to be just
below the unexplored millimetre region, once $n=2$ and $M_S \sim M_{ew}$.
Apart from classical gravitational experiments~\cite{expgravity},  
LSG effects are also accessible  via the effects of
spin--two gravitons that couple with SM 
particles and contribute to  pair production of bosons and
fermions  in $\rm e^+e^-$ collisions~\cite{hewett,giudice,agashe}, 
as described in terms of the parameter $M_S$~\cite{hewett},
interpreted as a cutoff of the theory. It appears as $1/M_S^4$ in the
LSG and SM interference terms and as $1/M_S^8$ in the pure graviton
exchange process~\cite{giudice,agashe}. These terms are 
multiplied by the  factors $\lambda$ and $\lambda^2$, respectively, which
incorporate the dependence on the unknown full LSG
theory and are of order unity~\cite{hewett}. To allow for both the
possible signs of the 
interference between the SM and LSG contributions the two cases
$\lambda = \pm 1$ are investigated. Figure~\ref{fig:5} presents
the modification to the
 $\epem\rightarrow\rm WW\rightarrow q\bar{q}' \ell \nu$,
  $\epem\rightarrow\gamma\gamma$ and 
  $\epem\rightarrow\mu^+\mu^-$
differential distributions and  $\rm e^+e^- \rightarrow e^+e^-$
differential cross sections in presence of LSG. 

\begin{table}[th]
  \begin{center}
    \begin{tabular}{|c|c|c|}
       \hline
       Process               &$M_S$(TeV)&$M_S$(TeV) \\
                             &$\lambda = +1$ & $\lambda = -1$ \\
       \hline
       $\rm e^+e^-\rightarrow ZZ$         & 0.77 & 0.76 \\
       $\rm e^+e^-\rightarrow W^+W^-$        & 0.79 & 0.68 \\
       $\rm e^+e^-\rightarrow\gamma\gamma$   & 0.79 & 0.80 \\
       \hline                  
       Bosons Combined         & 0.89 & 0.82 \\
       \hline                  
       $\rm e^+e^-\rightarrow\mu^+\mu^-$     & 0.69 & 0.56 \\
       $\rm e^+e^-\rightarrow\tau^+\tau^-$  & 0.54 & 0.58 \\
       $\rm e^+e^-\rightarrow q\bar{q}    $     & 0.49 & 0.49 \\
       $\rm e^+e^-\rightarrow e^+e^-     $     & 0.98 & 0.84 \\
       \hline                                    
       Fermions Combined          & 1.00 & 0.84 \\
       \hline                                    
       Bosons + Fermions         & 1.07 & 0.87 \\
       \hline
    \end{tabular}
    \caption{Lower limits at 95\% CL on the cutoff
    $M_S$ for different processes and values of
    $\lambda$}  
  \end{center}
\end{table}

From the analysis~\cite{l3lsg183,l3lsg189} of the presented distributions
as well as hadronic
W pair events and
$\tau$ pairs, together with the discriminating variables of the ZZ
event selections and the hadronic cross section, no
statistically significant hints for LSG are found in the L3 data at
$183\GeV$ and $189\GeV$ and the 95\% CL limits presented in Table~4 are set on $M_S$.
Assuming that no  higher order operators give sizeable
contributions to the LSG mediated boson and fermion pair production
and  that the meaning of the
cutoff parameter is the same for all the investigated processes, it is
possible to combine the boson and fermion limits. They are as high as
$1.07\TeV$ for $\lambda = +1$ and
$0.87\TeV$ for $\lambda = -1$ at 95\% CL and are competitive with
those achieved from a combined analysis of the LEP\,II data~\cite{melesanchez}.

The L3 collaboration also investigated the direct production of a
graviton, lost in the extra dimensions, 
associated with a
photon~\cite{l3singlephoton,l3lsg183} or a Z~\cite{l3lsg189}. The phase space favours the first
process. These searches did not not yield any evidence for the
expected LSG signatures~\cite{giudice,cheung} allowing to set 95\%  CL
limits on the LSG scale in excess of  $1\TeV$.

%
\section{\bf Conclusions and Omissions}
%

This walk through part of the recent L3 results shows a quite
wide activity, even on subjects that were not expected to be covered
at the start of the LEP\,II experimental program~\cite{yb}. On this
point all the experiments match the efforts of the accelerator
crew that is impressing the community with record--breaking
performance in term of energy, luminosity and operation efficiency.

Several important results have been omitted from this review due to
the lack of space and the framework of the discussion. The most
important worth mentioning are the 95\% CL mass limits on the SM Higgs
boson at $95.3\GeV$~\cite{h189} and that on the lightest MSSM neutralino at $32.5\GeV$~\cite{susy},
both achieved with the analysis of the full data sample collected by
L3 up to $189\GeV$. Moreover the L3 experiment has contributed with a
large amount of results to the field of two photon physics, as
reviewed in Reference~\cite{mw}.

Let me close this review with the hope that something
more than precise cross section measurements and interesting limits is
awaiting  the LEP community in the secrets of the last GeV still to be
squeezed from the machine in the high energy runs of 1999 up to a
centre--of--mass energy of $202\GeV$, and even beyond,  in the final run
of the year 2000.

%
\section*{\bf Acknowledgements}
%

I am grateful to  my colleagues of the L3 experiment, too numerous to be named here,
with whom I shared this challenging physics adventure since I arrived
at CERN as an undergraduate student, through the end of the century. It is only with the
lively daily interaction with many of them on several of the subjects I
described in this work, that I learned something
about experimental High Energy Physics.

I wish to thank the organisers of this conference in particular for
having helped me 
to find a swimming pool on the Leninskii prospekt and to fix some complex part of my travel
arrangements, two tasks at more than 95\% CL beyond the reach of my
Russian phrase book.


%

\end{document}